\documentclass[structabstract]{aa}  
\newcommand{\GILDAS}{\texttt{GILDAS}}
\newcommand{\CLASS}{\texttt{CLASS}}
\newcommand{\CLIC}{\texttt{CLIC}}
\newcommand{\MAPPING}{\texttt{MAPPING}}
\usepackage{color}
\usepackage{natbib}
\usepackage{amsmath}
\usepackage{graphicx}
\usepackage{txfonts}
\usepackage{subfig}
\usepackage{multirow} 
\usepackage{graphicx}
\usepackage{threeparttable}
\usepackage{sidecap}
\usepackage{booktabs}
\newcommand{\ie}{\emph{i.e.}}
\newcommand{\eg}{e.g.}
\newcommand{\emm}[1]{\ensuremath{#1}}   
\newcommand{\emr}[1]{\emm{\mathrm{#1}}} 
\newcommand{\Tkin}{\emm{T_\emr{kin}}}
\newcommand{\Tex}{\emm{T_\emr{ex}}}

\newcommand{\Td}{\emm{T_\emr{dust}}}
\newcommand{\nH}{\emm{n_\emr{H}}}
\newcommand{\dcop}{\emr{DCO^{+}}}                  
\newcommand{\h}{\emr{H}}                          
\newcommand{\x}{\emr{X}}                          
\newcommand{\hhco}{\emr{H_2CO}}            
\newcommand{\chhho}{\text{CH}_3\text{O}}            
\newcommand{\ohhco}{\text{o-H}_2\text{CO}}         
\newcommand{\phhco}{\emr{p-H_2CO}}         
\newcommand{\chhhoh}{\emr{CH_3OH}}         
\newcommand{\chhhohE}{\emr{CH_3OH-E}}         
\newcommand{\chhhohA}{\emr{CH_3OH-A}}         
\newcommand{\hh}{\emr{H_2}}                       
\newcommand{\ohh}{\text{o-H}_2}                    
\newcommand{\phh}{\text{p-H}_2}                    
\newcommand{\unit}[1]{\emm{\, \emr{#1}}}
\newcommand{\mm}{\unit{mm}}
\renewcommand{\deg}{\emm{^\circ}}
\newcommand{\pccm}{~\rm{cm}^{-3}}
\newcommand{\pscm}{~\rm{cm}^{-2}}
\newcommand{\ps}{~\rm{s}^{-1}}
\newcommand{\kms}{\emr{\,km\,s^{-1}}}

\newcommand{\mKkms}{~\rm{mK\,km\,s}^{-1}}
\newcommand{\Tsys}{\emm{T_\emr{sys}}}
\newcommand{\Tas}{\emm{T_\emr{A}^*}}
\newcommand{\Tmb}{\emm{T_\emr{mb}}}
\newcommand{\Beff}{\emm{B_\emr{eff}}}
\newcommand{\Feff}{\emm{F_\emr{eff}}}

\begin{document}
  \title{The IRAM-30m line survey of the Horsehead PDR:\\
     IV. Comparative chemistry of \hhco{} and \chhhoh{}\thanks{Based
       on observations obtained with the IRAM Plateau de Bure
       interferometer and 30~m telescope. IRAM is supported by
       INSU/CNRS (France), MPG (Germany), and IGN (Spain).}} 


   \author{V.V. Guzm\'{a}n\inst{1} \and J.R. Goicoechea\inst{2} \and
     J. Pety\inst{1,3} \and P. Gratier\inst{1} \and  M. Gerin\inst{3} \and
     E. Roueff\inst{4} \and F. Le Petit\inst{4} \and \\J. Le
     Bourlot\inst{4} \and A. Faure\inst{5}
         }

          \institute{Institut de Radioastronomie Millim\'etrique
            (IRAM), 300 rue de la Piscine, 38406 Saint Martin
            d'H\`eres, France\\  
            \email{[guzman,pety,gratier]@iram.fr}
            \and
            Departamento de Astrof\'{i}sica. Centro de Astrobiolog\'{i}a. 
            CSIC-INTA. Carretera de Ajalvir, Km 4. Torrej\'{o}n de Ardoz, \\
            28850 Madrid, Spain. \\
            \email{jr.goicoechea@cab.inta-csic.es}
            \and
            LERMA - LRA, UMR 8112, Observatoire de Paris and Ecole 
            normale Sup\'{e}rieure, 24 rue Lhomond, 75231 Paris, France. \\
            \email{maryvonne.gerin@lra.ens.fr}          
            \and
            LUTH UMR 8102, CNRS and Observatoire de Paris, Place J. Janssen, 
            92195 Meudon Cedex, France.\\
            \email{[evelyne.roueff,franck.lepetit,jacques.lebourlot]@obspm.fr}
            \and 
            UJF-Grenoble 1/CNRS-INSU, Institut de Plan\'etologie et
            d'Astrophysique de Grenoble (IPAG) UMR 5274, 38041
            Grenoble, France \\  
            \email{alexandre.faure@obs.ujf-grenoble.fr}
          }

   \date{Received August 7, 2013; accepted October 21, 2013}

 
  \abstract
   {Theoretical models and laboratory experiments show that \chhhoh{}
     is efficiently formed on cold grain surfaces through the
     successive hydrogenation of CO, forming HCO and \hhco{} as
     intermediate species. In cold cores and low UV-field illumination
     photo-dissociation regions (PDRs) the ices can be released into
     the gas-phase through non-thermal processes, like photodesorption,
     increasing considerably their gas-phase abundances.}
   {We investigate the dominant formation mechanism of \hhco{} and
     \chhhoh{} in the Horsehead PDR and its associated dense core.}
   {We performed deep integrations of several \hhco{} and \chhhoh{}
     lines at two positions in the Horsehead, namely the PDR and dense
     core, with the IRAM-30m telescope. In addition, we observed one
     \hhco{} higher frequency line with the CSO telescope at both
     positions. We determine the \hhco{} and \chhhoh{} column
     densities and abundances from the single-dish observations
     complemented with IRAM-PdBI high-angular resolution maps ($6''$)
     of both species. We compare the observed abundances with PDR
     models including either pure gas-phase chemistry or both
     gas-phase and grain surface chemistry.}
   {We derive \chhhoh{} abundances relative to total number of
     hydrogen atoms of $\sim1.2\times10^{-10}$ and
     $\sim2.3\times10^{-10}$ in the PDR and dense core positions,
     respectively. These abundances are similar to the inferred
     \hhco{} abundance in both positions ($\sim2\times10^{-10}$). We
     find an abundance ratio \hhco{}/\chhhoh{} of $\sim2$ in the PDR
     and $\sim1$ in the dense core. Pure gas-phase models cannot
     reproduce the observed abundances of either \hhco{} or \chhhoh{}
     at the PDR position. Both species are therefore formed on the
     surface of dust grains and are subsequently photodesorbed into
     the gas-phase at this position. At the dense core, on the other
     hand, photodesorption of ices is needed to explain the observed
     abundance of \chhhoh{}, while a pure gas-phase model can
     reproduce the observed \hhco{} abundance. The high-resolution
     observations show that \chhhoh{} is depleted onto grains at the
     dense core. \chhhoh{} is thus present in an envelope around this
     position, while \hhco{} is present in both the envelope and the
     dense core itself.}
   {Photodesorption is an efficient mechanism to release complex
     molecules in low FUV-illuminated PDRs, where thermal
     desorption of ice mantles is ineffective.}


    \maketitle
%

   \newcommand{\TabObsMaps}{%
     \begin{table*}
       \centering
       \caption{Observation parameters for the maps shown in
         Figs.~\ref{fig:maps30m} and \ref{fig:mapsPdBI}. The
         projection center of all the maps is $\alpha_{2000} =
         05^h40^m54.27^s$, $\delta_{2000} = -02\deg 28' 00''$.}
       \begin{threeparttable}
         {\tiny
           \begin{tabular}{ccrcccccccl}\toprule
             Molecule & Transition & Frequency  & Instrument & Mode & Beam
             & Vel. Resol. & Int. Time & \Tsys{} & Noise & Reference\\
             & & GHz & & & arcsec & $\kms$ & hours & K (\Tas{}) & K (\Tmb{}) &\\
             \midrule
             \multicolumn{3}{c}{Continuum at 1.2\mm} & 30m/MAMBO  & --- & $11.7\times11.7$ &  --  & -- &  -- & -- & \cite{hilyblant05}\\
             \dcop{}    & $3-2$           & 216.112582 & 30m/HERA & FSW & $11.4\times11.4$ & 0.11 & 1.5/2.0$^{a}$ & 230 & 0.10      & \cite{pety07}\\
             $\phhco$   & $3_{03}-2_{02}$ & 218.222190 & 30m/HERA & PSW & $11.9\times11.9$ & 0.05  & 2.1/3.4$^a$  & 280 & 0.32      & \cite{guzman11}\\
             $\phhco$   & $2_{02}-1_{01}$ & 145.602949 & 30m/EMIR & PSW & $17.8\times17.8$ & 0.20 & 7.4/12.9$^a$  & 208 & 0.17      & This work \\
             $\chhhohE$ & $3_{-1}-2_{-1}$ & 145.097370 & 30m/EMIR & PSW & $17.9\times17.9$ & 0.20 & 7.4/12.9$^a$  & 208 & 0.108$^b$ & This work \\
                        &                 &            & 30m/C150 & FSW & $17.9\times17.9$ & 0.20 &  2.6/3.2$^a$  & 163 & 0.108$^b$ & This work \\
             $\chhhohA$ & $3_{0}-2_{0}$   & 145.103152 & 30m/EMIR & PSW & $17.9\times17.9$ & 0.20 & 7.4/12.9$^a$  & 208 & 0.095$^b$ & This work \\
                        &                 &            & 30m/C150 & FSW & $17.9\times17.9$ & 0.20 &  2.6/3.2$^a$  & 263 & 0.095$^b$ & This work \\
             \bottomrule
         \end{tabular}}
         \begin{tablenotes}[para,flushleft]
           $^{a}$ Two values are given for the integration time: the
           on-source time
           and the telescope time.\\
           $^{b}$ The noise value is computed on the final maps that
           combines the data from both observing runs (2007 \& 2012).\\
         \end{tablenotes}
             {\tiny
               \begin{tabular}{ccrcccccccl}\toprule
                 Molecule & Transition & Frequency & Instrument & Beam & PA & Vel. Resol. & Int. Time & \Tsys{} & Noise & Reference\\
                 &            & GHz        &            & arcsec & $\deg$ & $\kms$     & hours     & K (\Tas{}) & mK (\Tmb{}) & \\
                 \midrule
                 HCO & $1_{01}\,3/2,2-0_{00}\,1/2,1$ & 86.670760 & PdBI/C\&D & $6.7 \times 4.4$ & 16 & 0.20 & 6.5$^{b}$ & 150 & 90$^{c}$ & \cite{gerin09} \\
                 $\phhco$ & $2_{02}-1_{01}$   & 145.602949 & PdBI/C\&D & $6.1\times5.6$ & 166 & 0.20 & 5.3/19$^{b}$ & 145 & 244 & This work \\
                 $\chhhohE$ & $3_{-1}-2_{-1}$ & 145.097370 & PdBI/C\&D & $6.1\times5.6$ & 166 & 0.20 & 5.3/19$^{b}$ & 145 & 116 & This work \\
                 $\chhhohA$ & $3_{0}-2_{0}$   & 145.103152 & PdBI/C\&D & $6.1\times5.6$ & 166 & 0.20 & 5.3/19$^{b}$ & 145 & 127 & This work \\
                 \bottomrule
             \end{tabular}}
             \begin{tablenotes}[para,flushleft]
               $^{b}$ Two values are given for the integration time: the
               on-source time (as if the source were always observed with 6
               antennae) and the telescope time.\\
               $^{c}$ The noise values quoted here are the noises at the mosaic phase 
               center (Mosaic noise is inhomogeneous due to primary beam correction; it 
               steeply increases at the mosaic edges).
             \end{tablenotes}
       \end{threeparttable}
       \label{tab:obs:maps} 
     \end{table*}
   }

   \newcommand{\TabObsLines}{%
     \begin{table*}
       \centering
       \caption{Observation parameters of the deep integrations of the
         \chhhoh{} lines detected with the 30m and the \hhco{} line
         detected with the CSO towards the PDR and the dense core.}
       \begin{threeparttable}
         \begin{tabular}{lllrrccrrr}\toprule
           Position & Molecule & Transition & $\nu$ & Line area & Velocity & FWHM &
           $T_{\textrm{peak}}$ & RMS & S/N \\
           & & & GHz & $\mKkms$ & $\kms$ & $\kms$ & mK & mK &\\
           \midrule
           \multirow{9}{*}{PDR} %
           & $\chhhohE$ & $5_{-1}-4_{0}$  &  84.521 &  29.1$\pm$ 3.1 & 10.64$\pm$0.05 & 0.98$\pm$0.14 &  27.9 &   4.7 &   6\\ 
           & $\chhhohE$ & $2_{-1}-1_{-1}$ &  96.739 &  74.1$\pm$ 4.3 & 10.73$\pm$0.02 & 0.56$\pm$0.04 & 125.0 &   9.4 &  13\\ 
           & $\chhhohE$ & $2_{0}- 1_{0}$  &  96.745 &  22.3$\pm$ 4.7 & 10.68$\pm$0.05 & 0.63$\pm$0.21 &  33.2 &   7.5 &   4\\ 
           & $\chhhohE$ & $3_{0}- 2_{0}$  & 145.094 &  26.9$\pm$ 6.5 & 10.77$\pm$0.05 & 0.42$\pm$0.11 &  59.8 &  21.7 &   3\\ 
           & $\chhhohE$ & $3_{-1}-2_{-1}$ & 145.097 &  98.4$\pm$ 9.0 & 10.54$\pm$0.02 & 0.51$\pm$0.06 & 180.8 &  21.5 &   8\\ 
           & $\chhhohE$ & $5_{-1}-4_{-1}$ & 241.767 &  81.0$\pm$ 8.8 & 10.66$\pm$0.04 & 0.85$\pm$0.11 &  89.5 &  13.1 &   7\\ 
           & $\chhhohA$ & $2_{0}- 1_{0}$  &  96.741 & 120.3$\pm$ 3.9 & 10.73$\pm$0.01 & 0.61$\pm$0.02 & 186.6 &   7.7 &  24\\ 
           & $\chhhohA$ & $3_{0}- 2_{0}$  & 145.103 & 145.7$\pm$ 7.0 & 10.66$\pm$0.01 & 0.51$\pm$0.03 & 266.2 &  19.4 &  14\\ 
           & $\chhhohA$ & $5_{0}- 4_{0}$  & 241.791 &  59.5$\pm$ 7.3 & 10.81$\pm$0.03 & 0.48$\pm$0.08 & 117.1 &  12.5 &   9\\ 
           & $\ohhco$   & $4_{13}-3_{12}$ & 300.837 & 204.1$\pm$39.0 & 10.75$\pm$0.07 & 0.68$\pm$0.16 & 281.1 & 60.8 & 5\\
           \midrule                                                          
           \multirow{16}{*}{Core} %
           & $\chhhohE$ & $5_{-1}-4_{0}$  &  84.521 &  43.8$\pm$ 3.0 & 10.64$\pm$0.02 & 0.65$\pm$0.05 &  63.3 &  6.4 &  10\\ 
           & $\chhhohE$ & $2_{-1}-1_{-1}$ &  96.739 & 304.4$\pm$ 3.5 & 10.65$\pm$0.00 & 0.57$\pm$0.01 & 497.5 &  7.9 &  63\\ 
           & $\chhhohE$ & $2_{0}- 1_{0}$  &  96.745 &  57.6$\pm$ 2.7 & 10.66$\pm$0.01 & 0.50$\pm$0.03 & 107.6 &  5.7 &  19\\ 
           & $\chhhohE$ & $2_{1}- 1_{1}$  & 96.755 &  $11.0\pm3.0$ & $10.56\pm0.05$ & $0.41\pm0.14$ & 25.0 & 7.0 &  3\\
           & $\chhhohE$ & $0_{0}- 1_{-1}$ & 108.894 &  74.4$\pm$ 4.6 & 10.63$\pm$0.01 & 0.46$\pm$0.03 & 151.8 & 13.5 &  11\\ 
           & $\chhhohE$ & $3_{0}- 2_{0}$  & 145.094 &  50.4$\pm$ 6.9 & 10.48$\pm$0.02 & 0.31$\pm$0.05 & 154.5 & 24.3 &   6\\ 
           & $\chhhohE$ & $3_{-1}-2_{-1}$ & 145.097 & 449.7$\pm$ 7.5 & 10.48$\pm$0.00 & 0.52$\pm$0.01 & 819.8 & 21.6 &  38\\ 
           & $\chhhohE$ & $3_{1}-2_{1}$ & 145.132 & $<12$ &- & $0.5^a$ &- & 22.0 &-\\
           & $\chhhohE$ & $1_{0}- 1_{-1}$ & 157.271 & 122.3$\pm$11.0 & 10.66$\pm$0.03 & 0.56$\pm$0.05 & 206.9 & 34.8 &   6\\ 
           & $\chhhohE$ & $2_{0}- 2_{-1}$ & 157.276 &  69.5$\pm$ 9.7 & 10.64$\pm$0.03 & 0.42$\pm$0.07 & 154.0 & 28.1 &   5\\ 
           & $\chhhohE$ & $1_{1}- 0_{0}$  & 213.427 &  42.8$\pm$ 6.4 & 10.64$\pm$0.04 & 0.46$\pm$0.07 &  87.5 & 12.2 &   7\\ 
           & $\chhhohE$ & $4_{2}- 3_{1}$  & 218.440 &  47.4$\pm$ 5.6 & 10.62$\pm$0.04 & 0.62$\pm$0.08 &  71.5 &  9.2 &   8\\ 
           & $\chhhohE$ & $5_{-1}-4_{-1}$ & 241.767 & 170.8$\pm$ 7.9 & 10.57$\pm$0.01 & 0.52$\pm$0.03 & 311.5 & 15.3 &  20\\ 
           & $\chhhohE$ & $2_{0}- 1_{-1}$ & 254.015 &  59.6$\pm$ 4.3 & 10.53$\pm$0.02 & 0.44$\pm$0.03 & 126.5 &  7.9 &  16\\ 
           & $\chhhohE$ & $2_{1}- 1_{0}$  & 261.806 &  43.3$\pm$ 7.3 & 10.56$\pm$0.03 & 0.29$\pm$0.14 & 141.1 & 22.5 &   6\\ 
           & $\chhhohA$ & $2_{0}- 1_{0}$  &  96.741 & 448.5$\pm$ 3.9 & 10.65$\pm$0.00 & 0.58$\pm$0.01 & 725.6 &  8.1 &  89\\ 
           & $\chhhohA$ & $3_{0}- 2_{0}$  & 145.103 & 527.2$\pm$ 8.0 & 10.56$\pm$0.00 & 0.52$\pm$0.01 & 949.5 & 21.3 &  45\\ 
           & $\chhhohA$ & $5_{0}- 4_{0}$  & 241.791 & 200.2$\pm$ 8.9 & 10.67$\pm$0.01 & 0.47$\pm$0.02 & 398.4 & 16.2 &  25\\ 
           & $\ohhco$   & $4_{13}-3_{12}$   & 300.837 & 211.1$\pm$33.0 & 10.78$\pm$0.05 & 0.62$\pm$0.10 & 320.8 & 56.4 & 6\\
          \bottomrule
         \end{tabular}
         \begin{tablenotes}[para,flushleft]
           $^a$ Fixed to compute the upper limit in line area.\\
           Note: All temperatures are given in the main
           beam temperature scale.
         \end{tablenotes}
       \end{threeparttable}
       \label{tab:obs:lines}
     \end{table*}
   }

\newcommand{\SpecParam}{%
\begin{table}
  \caption{Spectroscopic parameters of the detected lines obtained from the CDMS data base \citep{muller01}. }
  \begin{center}
  \begin{tabular}{rcrrcrccc}\toprule
    Molecule & Transition & $\nu$ & $E_u$ & $A_{ul}$ & $g_u$ \\
    & & [GHz] & [K] & [s$^{-1}$] & \\
   \midrule
   $\chhhohE$ & $5_{-1}-4_{0}$   &  84.521 & 40.4 & $2.0\times10^{-6}$ & 11\\
   $\chhhohE$ & $2_{-1}-1_{-1}$  &  96.739 & 12.5 & $3.0\times10^{-6}$ & 5\\
   $\chhhohE$ & $2_{0}- 1_{0}$   &  96.745 & 20.1 & $3.0\times10^{-6}$ & 5\\
   $\chhhohE$ & $0_{0}- 1_{-1}$  & 108.894 & 13.1 & $1.5\times10^{-5}$ & 1\\
   $\chhhohE$ & $3_{0}- 2_{0}$   & 145.094 & 27.1 & $1.2\times10^{-5}$ & 7\\
   $\chhhohE$ & $3_{-1}-2_{-1}$  & 145.097 & 19.5 & $1.1\times10^{-5}$ & 7\\
   $\chhhohE$ & $1_{0}- 1_{-1}$  & 157.271 & 15.4 & $2.2\times10^{-5}$ & 3\\
   $\chhhohE$ & $2_{0}- 2_{-1}$  & 157.276 & 20.1 & $2.2\times10^{-5}$ & 5\\
   $\chhhohE$ & $1_{1}- 0_{0}$   & 213.427 & 23.4 & $3.4\times10^{-5}$ & 3\\
   $\chhhohE$ & $4_{2}- 3_{1}$   & 218.440 & 45.5 & $4.7\times10^{-5}$ & 9\\
   $\chhhohE$ & $5_{-1}-4_{-1}$  & 241.767 & 40.4 & $5.8\times10^{-5}$ & 11\\
   $\chhhohE$ & $2_{0}- 1_{-1}$  & 254.015 & 20.1 & $1.9\times10^{-5}$ & 5\\
   $\chhhohE$ & $2_{1}- 1_{0}$   & 261.806 & 28.0 & $5.6\times10^{-5}$ & 5\\
   \midrule
   $\chhhohA$ & $2_{0}- 1_{0}$   &  96.741 &  7.0 & $3.0\times10^{-6}$ & 5\\
   $\chhhohA$ & $3_{0}- 2_{0}$   & 145.103 & 13.9 & $1.2\times10^{-5}$ & 7\\
   $\chhhohA$ & $5_{0}- 4_{0}$   & 241.791 & 34.8 & $6.0\times10^{-5}$ & 11\\
   \bottomrule
  \end{tabular}
  \end{center}
  \label{tab:spec_param}
\end{table}
}

\newcommand{\TabCritDens}{%
\begin{table}[t!]
  \caption{Critical densities$^a$ ($\pccm$) for the \chhhoh{} lines
    detected in this work with $\phh$ and $\ohh$ as colliding partners
    computed for two kinetic temperatures.}  \centering
  \begin{threeparttable}[b]
  \begin{tabular}{cccccc}\toprule
    & & \multicolumn{2}{c}{$\Tkin=20$~K} & \multicolumn{2}{c}{$\Tkin=60$~K} \\
    Type & $J_{K}$ & $\phh$    &   $\ohh$     & $\phh$    &  $\ohh$\\         
    \midrule
    \multirow{10}{*}{E} 
    & $5_{-1}$ & $2.47\times10^5$ & $1.91\times10^5$ & $2.52\times10^5$ & $2.08\times10^5$\\ 
    & $2_{-1}$ & $2.81\times10^4$ & $2.33\times10^4$ & $2.66\times10^4$ & $2.33\times10^4$\\
    & $2_{ 0}$ & $3.28\times10^5$ & $1.52\times10^5$ & $3.59\times10^5$ & $1.55\times10^5$\\
    & $0_{ 0}$ & -                & $1.58\times10^5$ & -                & $2.16\times10^5$\\
    & $3_{ 0}$ & $4.13\times10^5$ & $2.88\times10^5$ & $4.40\times10^5$ & $2.83\times10^5$\\
    & $3_{-1}$ & $7.05\times10^4$ & $6.88\times10^4$ & $6.88\times10^4$ & $6.83\times10^4$\\
    & $1_{ 0}$ & $3.73\times10^5$ & $1.11\times10^5$ & $3.75\times10^5$ & $1.14\times10^5$\\
    & $1_{ 1}$ & $2.70\times10^6$ & $2.62\times10^5$ & $3.83\times10^6$ & $2.83\times10^5$\\
    & $4_{ 2}$ & $3.58\times10^5$ & $1.87\times10^5$ & $3.50\times10^5$ & $1.94\times10^5$\\
    & $2_{ 1}$ & $6.75\times10^5$ & $2.88\times10^5$ & $7.19\times10^5$ & $2.83\times10^5$\\
    \midrule
    \multirow{3}{*}{A} 
    & $5_{-1}$ & $2.28\times10^5$ & $2.06\times10^5$ & $2.45\times10^5$ & $2.20\times10^5$\\
    & $2_{-1}$ & $6.70\times10^4$ & $6.56\times10^4$ & $7.04\times10^4$ & $7.35\times10^4$\\
    & $2_{ 0}$ & $2.86\times10^4$ & $2.29\times10^4$ & $2.96\times10^4$ & $2.58\times10^4$\\
    \bottomrule
  \end{tabular}
  \begin{tablenotes}[para,flushleft]
    $^a$ Computed as:~$n_{cr}(J'_{K'_a K'_c}, \Tkin) =
    \frac{\sum_{J''_{K''_a K''_c}} A( J'_{K'_a K'_c} \rightarrow
      J''_{K''_a K''_c})}{\sum_{J''_{K''_a K''_c}}
    \gamma(J'_{K'_a K'_c} \rightarrow
    J''_{K''_a K''_c}, \Tkin)}
  $
  \end{tablenotes}
  \end{threeparttable}
  \label{tab:ncrit}
\end{table}
}

\newcommand{\FigLines}{%
  \begin{figure*}[t!]
    \centering
    \includegraphics[scale=0.4]{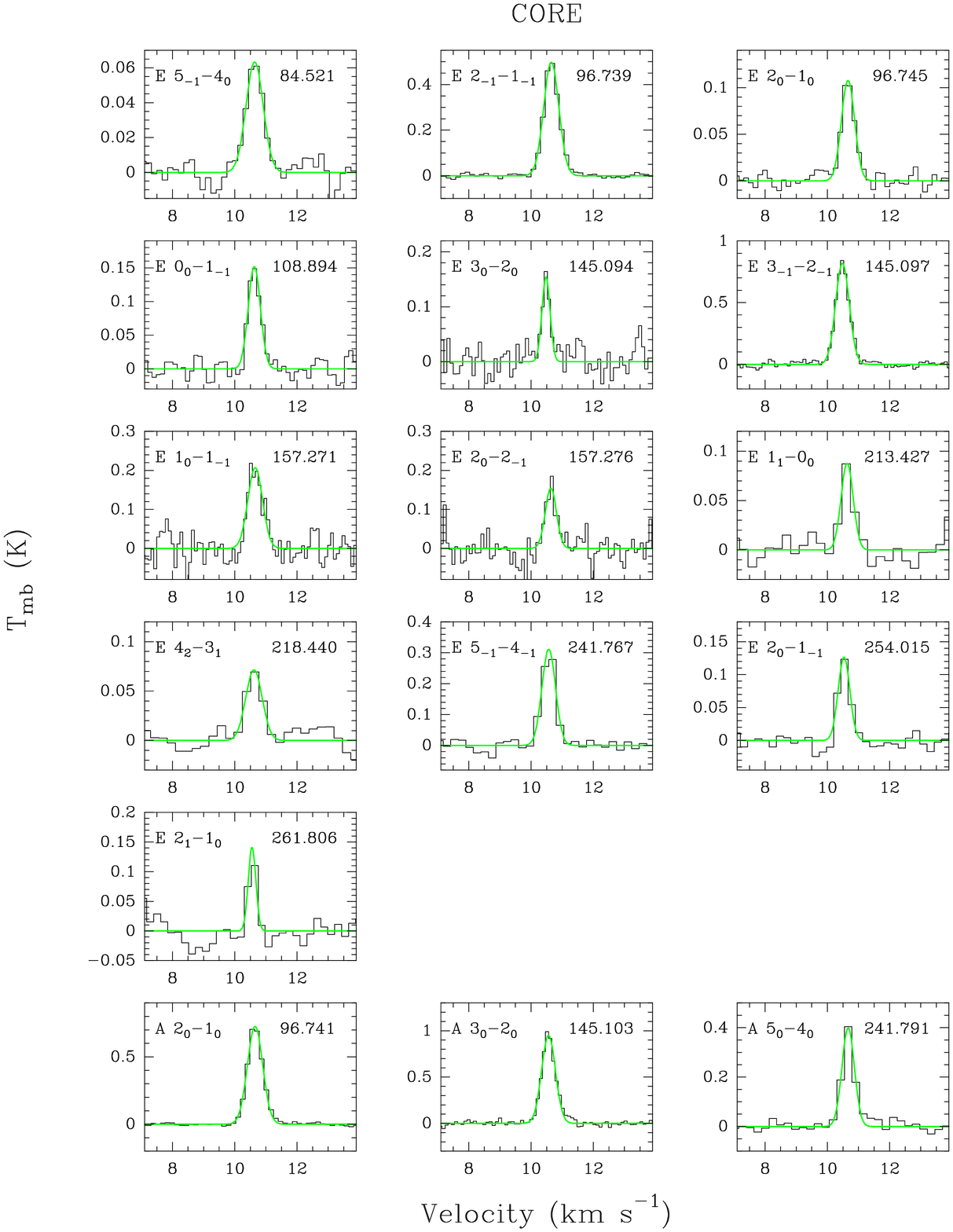} 
    \hfill
    \includegraphics[scale=0.4]{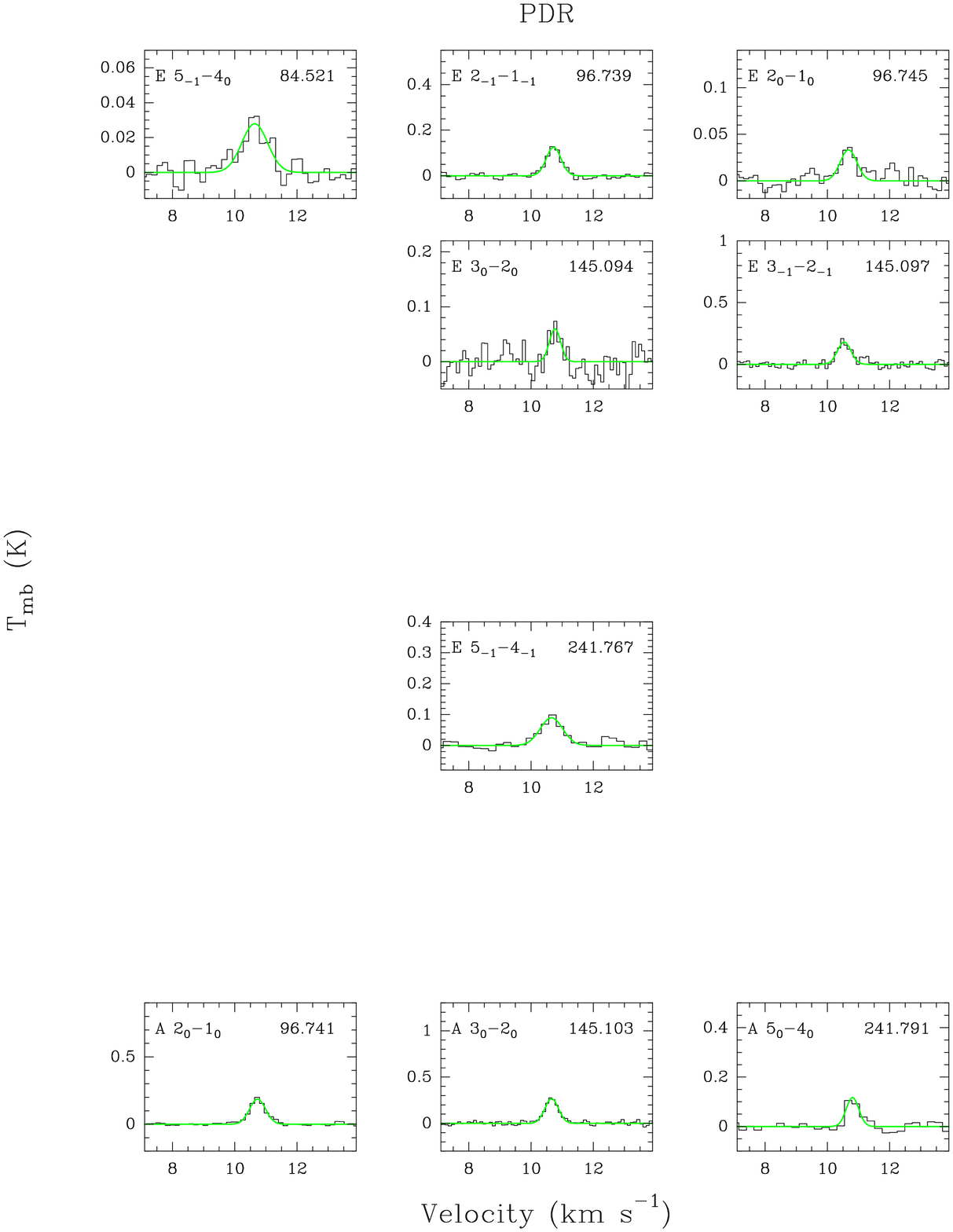} 
    \caption{Detected \chhhoh{} lines towards the dense core
      (\textit{left}) and PDR (\textit{right}) positions. The green
      curves are Gaussian fits. The line frequency in GHz is given in
      each box. For each line, the same scale is used at both
      positions to ease the comparison.}
    \label{fig:lines}
  \end{figure*}
}

\newcommand{\FigCSO}{%
  \begin{figure}[b!]
    \centering
    \includegraphics[scale=0.55]{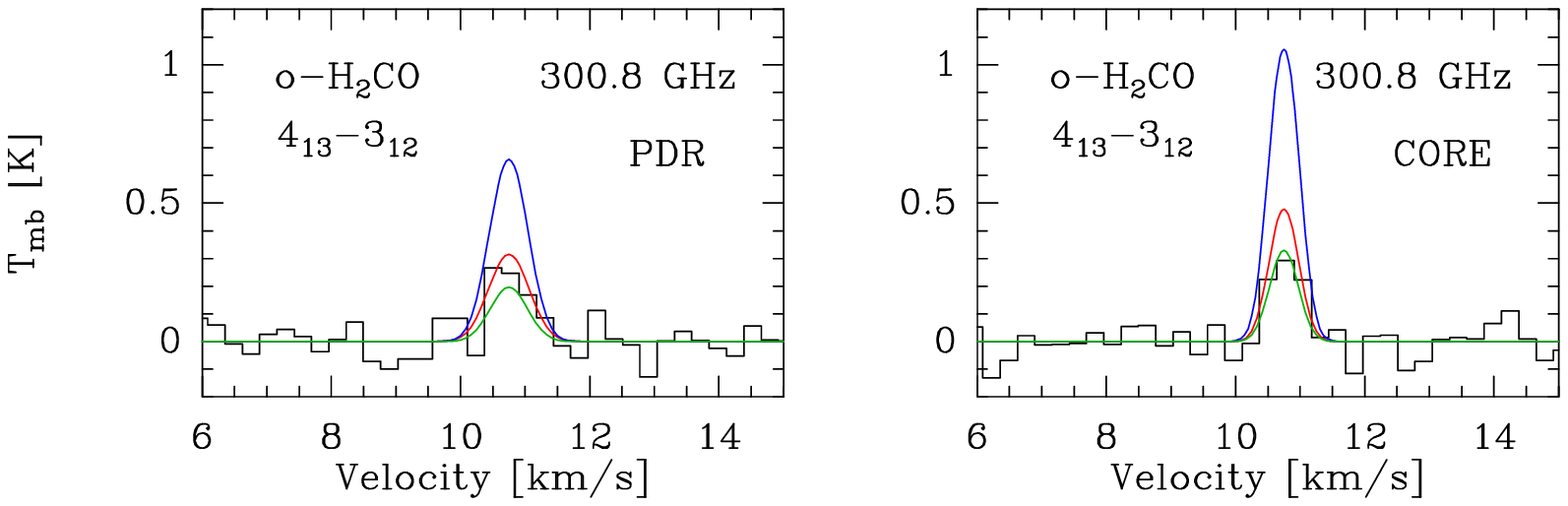} 
    \caption{New $\ohhco$ line detected towards the PDR
      (\textit{left}) and dense core (\textit{right}) positions. The
      three color lines are radiative transfer models from
      \cite{guzman11}, where the column density was varied around the
      best match (red curve) by a factor of 1.5 (blue curve) and 1/1.5
      (green curve). The column density used for the best fit is given
      in Table~\ref{tab:column_densities}. The line frequency in GHz is
      given in each box.}
    \label{fig:cso}
  \end{figure}
}

\newcommand{\FigMontecarlo}{%
  \begin{figure*}[t!]
    \centering
    \includegraphics[scale=.6]{figures/mtc_new_all-includeCSO.eps}
    \caption[Optional caption for list of figures]{Radiative transfer
      modeling of \hhco{} lines reported by \cite{guzman11} and the
      new detection reported in this work. Two left columns: the PDR
      position ($\Tkin=60$ K, $n(\hh)~=~6~\times~10^4 \pccm$,
      N($\ohhco$)~=~$7.2~\times~10^{12} \pscm$) and two right columns:
      the dense core position ( $\Tkin=20$ K, $n(\hh)~=~10^5 \pccm$,
      N($\ohhco$)~=~$9.6~\times~10^{12} \pscm$). The three top rows
      display the ortho lines, for which we varied the column density
      around the best match (red curve) by a factor of 1.5 (blue
      curve) and 1/1.5 (green curve). The two bottom rows display the
      para lines, for which we keep constant the column density of the
      best match for $\ohhco$ (red curves) and varied the
      ortho-to-para ratio of \hhco{}: o/p = 1.5 (dashed blue), o/p=2
      (dashed red) and o/p=3 (dashed green). See \cite{guzman11} for
      more details on the radiative transfer model.}
    \label{fig:montecarlo}
  \end{figure*}
}

\newcommand{\FigPDRmodelIces}{%
  \begin{figure*}[t!]
    \centering
    \includegraphics[scale=.79]{figures/abundances_conv-ices.eps}
    \caption[Optional caption for list of figures]{Gas and ice
      abundances predicted by the model.}
    \label{fig:ices}
  \end{figure*}
}

\newcommand{\FigFWHM}{%
  \begin{figure*}[t!]
    \centering
    \includegraphics[scale=.7]{figures/h2co-ch3oh-linewidths.eps}
    \caption{Line widths.}
    \label{fig:fwhm}
  \end{figure*}
}

\newcommand{\FigLevelDiag}{%
  \begin{figure}
    \begin{center}
      \includegraphics[scale=0.45]{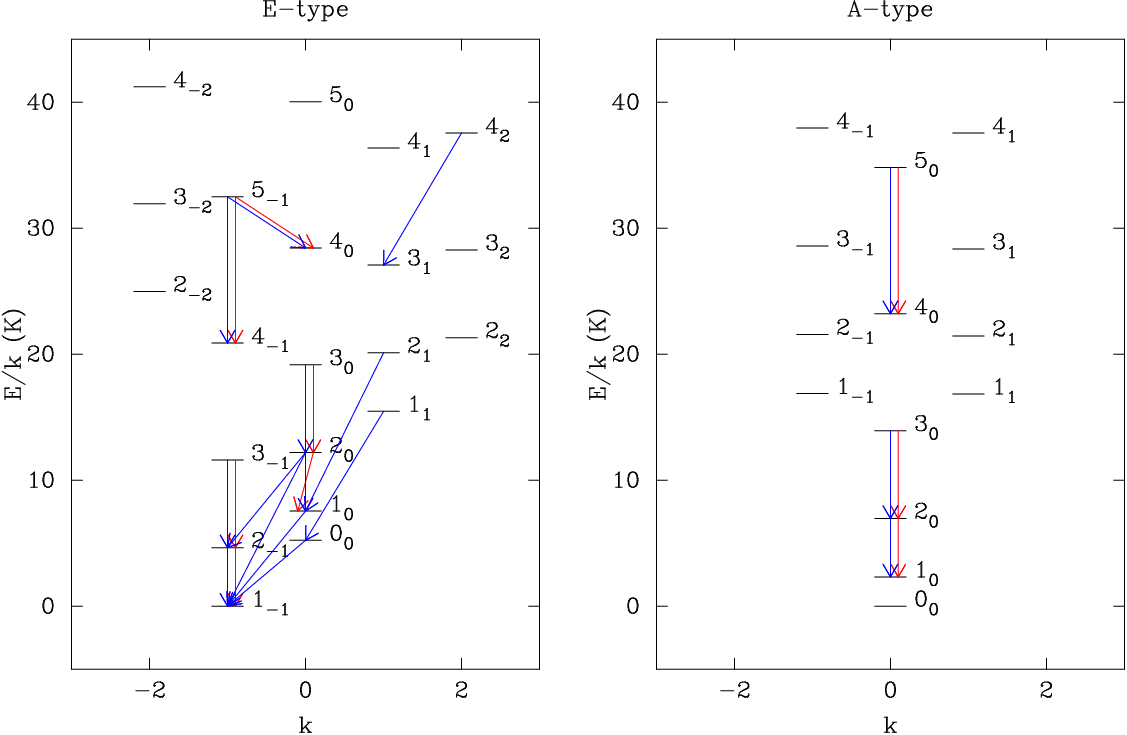}
    \end{center}
    \caption{Lower energy rotational levels of $\chhhohE$ (\textit{left}) and $\chhhohA$
      (\textit{right}). The arrows indicate the lines detected in the PDR
      (red) and the core (blue).}
    \label{fig:level_diag}
  \end{figure}
}

\newcommand{\FigRotDiag}{%
  \begin{figure}[h!]
    \hspace{-0.5cm}
    \includegraphics[scale=0.55]{rot-diag-ch3oh.eps}
    \caption{Rotational diagrams}
    \label{fig:rot-diag}
  \end{figure}
}

\newcommand{\FigChi}{%
  \begin{figure}[t!]
    \centering
    \includegraphics[scale=0.45]{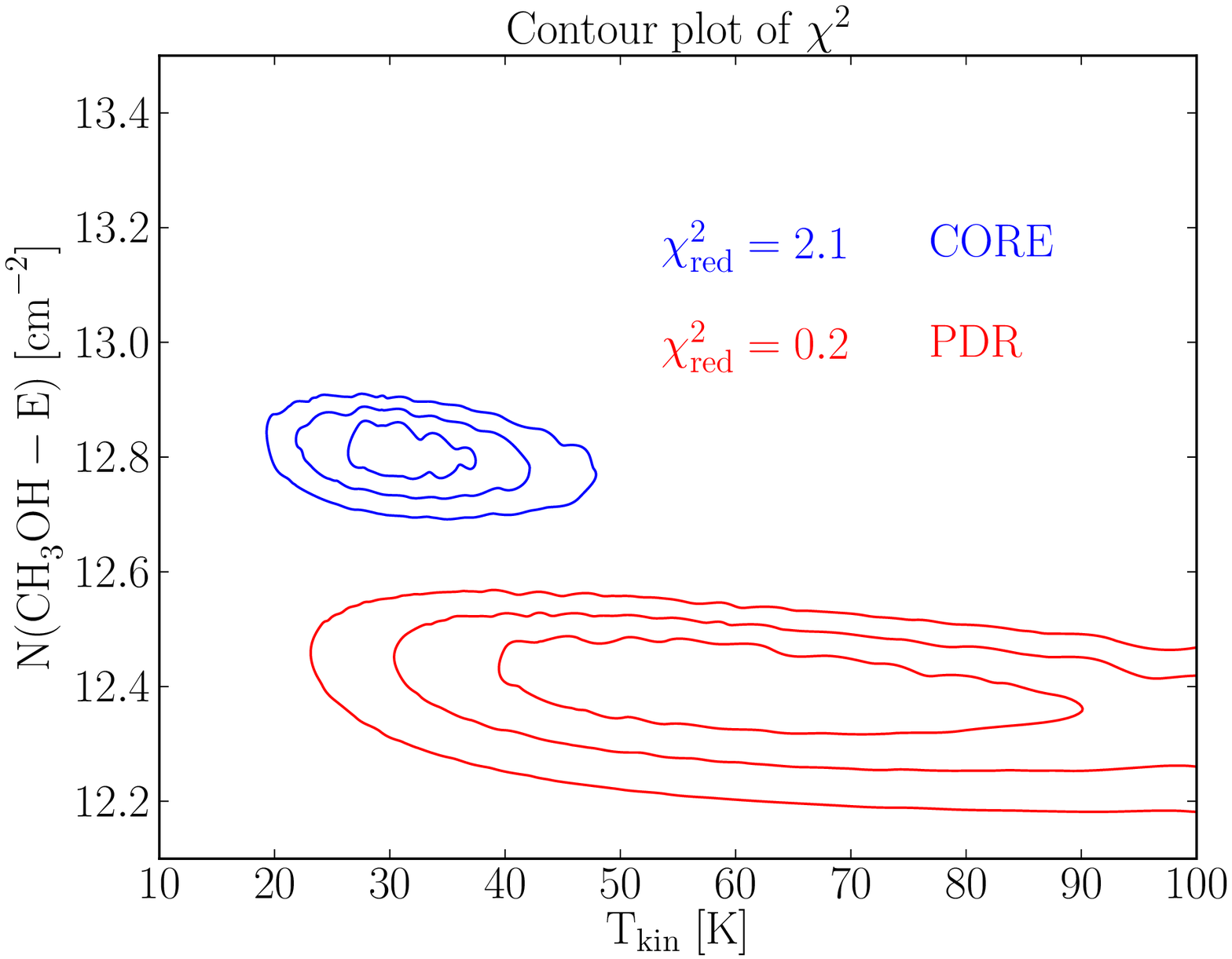}
    \caption{$\chi^2$ as a function of N($\chhhohE$) and $\Tkin$ for
      the PDR (red) and dense core (blue) positions. The $\hh$ density
      was kept constant to n(H$_2$) = $6\times10^{4} \pccm$ (PDR) and
      n(H$_2$) = $1\times10^{5} \pccm$ (dense core). The contours
      indicate the 1, 2 and 3$\sigma$ confidence levels for the
      models. The reduced $\chi^2$, defined as $\chi^2_{\mathrm{red}}
      = \chi^2/(N-2)$, is shown for the best fit model at each
      position.}
    \label{fig:chi2}
  \end{figure}
}

\newcommand{\TabColumnDens}{%
  \begin{table}[t!]
    \centering
    \caption{Column densities and abundances.}
    \begin{threeparttable}[b]
      \begin{tabular}{clcc}\toprule
        & Molecule             &   PDR             & Dense core         \\
        \midrule
        &$N_{\mathrm{H}}$ & $3.8\times10^{22}$ & $6.4\times10^{22}$ \\
        \multirow{2}{*}{Column}  &$N$(HCO)$^a$    & $3.2\times10^{13}$ & $<4.6\times10^{12}$\\
        \multirow{2}{*}{density}  &$N(\ohhco)^b$    & $7.2\times10^{12}$ & $9.6\times10^{12}$  \\
        \multirow{2}{*}{$[\pscm]$} &$N(\phhco)^b$    & $3.6\times10^{12}$ & $3.2\times10^{12}$  \\
        &$N(\chhhohE)$ & $2.7\times10^{12}$ & $6.5\times10^{12}$ \\ 
        &$N(\chhhohA)$ & $2.0\times10^{12}$ & $8.1\times10^{12}$ \\ 
        \midrule
        &[HCO]  & $8.4\times10^{-10}$ & $<7.2\times10^{-11}$ \\
        \multirow{2}{*}{Abundances}  &$[\ohhco]$     & $1.9\times10^{-10}$ & $1.5\times10^{-10}$ \\
        \multirow{2}{*}{$\frac{N(\x)}{(N(\h)+2~N(\hh))}$} &$[\phhco]$ & $9.5\times10^{-11}$ & $5.0\times10^{-11}$ \\
        &$[\chhhohE]$  & $7.0\times10^{-11}$ & $1.0\times10^{-10}$ \\
        &$[\chhhohA]$  & $5.3\times10^{-11}$ & $1.3\times10^{-10}$ \\
        \midrule
        &$\hhco / \chhhoh^c$ & $2.3\pm0.4$ & $0.9\pm0.1$\\
        \bottomrule
      \end{tabular}
      \begin{tablenotes}[para,flushleft]
        $^a$ \cite{gerin09}.\\
        $^b$ \cite{guzman11}.\\
        $^c$ We estimate an error of $\sim15$\% for the \hhco{}
        column density (similar to \chhhoh{}) to compute the error in
        the ratio.\\
      \end{tablenotes}
    \end{threeparttable}
    \label{tab:column_densities}
  \end{table}
}

\newcommand{\FigPDRmodel}{%
  \begin{figure*}
    \centering 
    \includegraphics[scale=0.79]{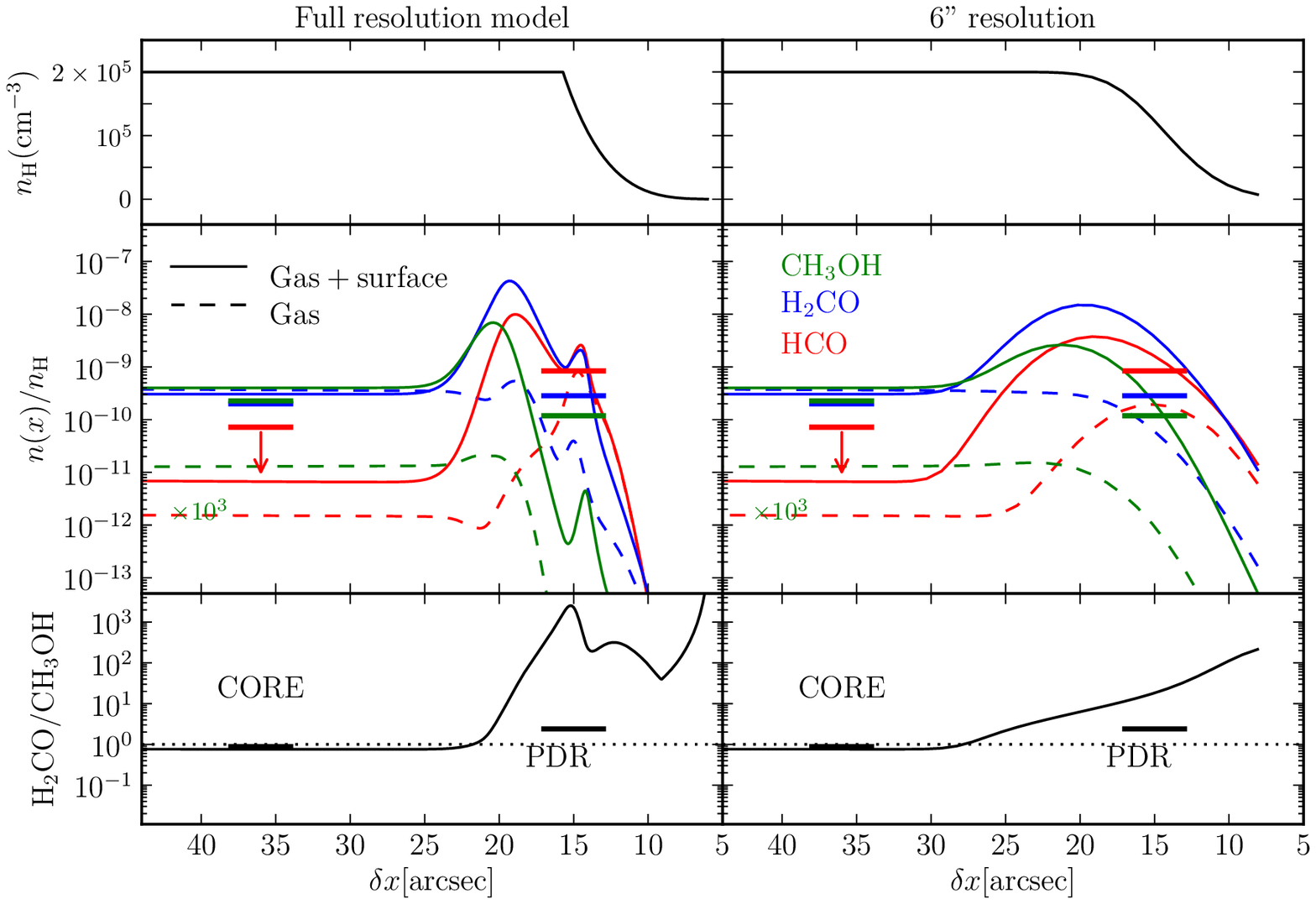}
    \caption{Photochemical model of the Horsehead PDR. The
      \textit{left column} shows the full resolution model and the
      \textit{right column} shows the model convolved with a
      Gaussian of 6'' FWHM. \textit{Upper panel:} PDR density
      profile ($\nH = n($H$) + 2 n($H$_2)$ in
      $\pccm$). \textit{Middle panel:} Predicted abundance (relative
      to $\nH$) of HCO (red), \hhco{} (blue) and \chhhoh{}
      (green). \textit{Lower panel:} Predicted $\hhco / \chhhoh$
      abundance ratio. Models shown as dashed lines include pure
      gas-phase chemistry and models shown as solid lines include
      both gas-phase and grain surface chemistry. The \chhhoh{}
      abundance predicted by the pure gas-phase model has been
      multiplied by $10^3$ in the middle panel so it appears in the
      figure.}
    \label{fig:meudon}
  \end{figure*}
}

\newcommand{\FigMapsSingleDish}{%
  \begin{figure*}[t!]
    \centering
    \includegraphics[scale=0.6]{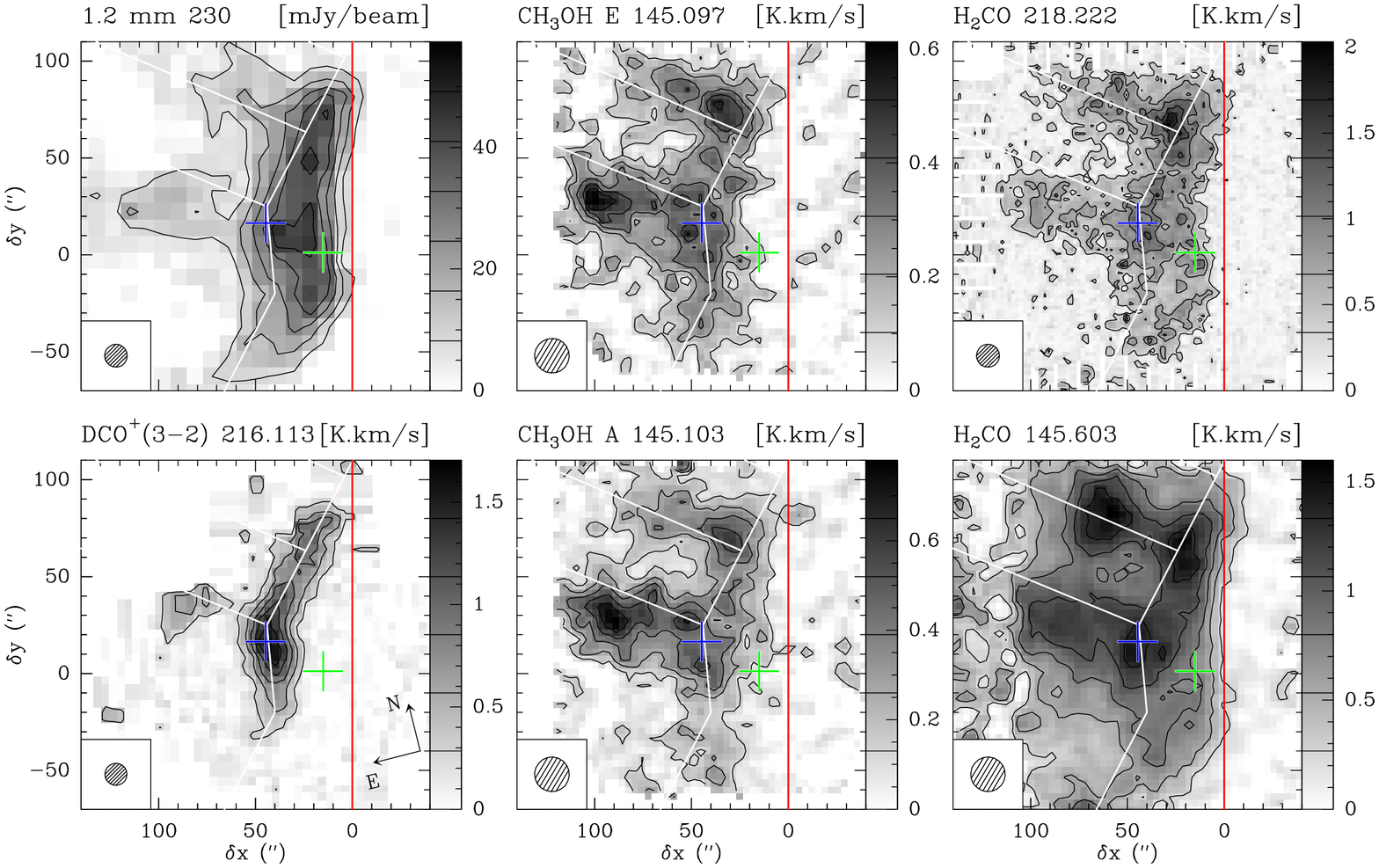}
    \caption{IRAM-30m maps of the Horsehead edge. Maps were rotated by
      $14^{\deg}$ counter–clockwise around the projection center,
      located at $(\delta x, \delta y) = (20''′, 0′'')$, to bring the
      exciting star direction in the horizontal direction and the
      horizontal zero was set at the PDR edge, delineated by the red
      vertical line. The crosses show the positions of the PDR (green)
      and the dense-core (blue), where deep integrations were
      performed at IRAM-30m. The white lines delineate the arc-like
      structure of the \dcop{} emission. The spatial resolution is
      plotted in the bottom left corner. Values of contour levels are
      shown on each image lookup table. The emission of all lines is
      integrated between 10.1 and 11.1 $\kms$.}
    \label{fig:maps30m}
  \end{figure*}
}

\newcommand{\FigMapsPdBI}{%
  \begin{figure}
    \centering
    \includegraphics[scale=0.5]{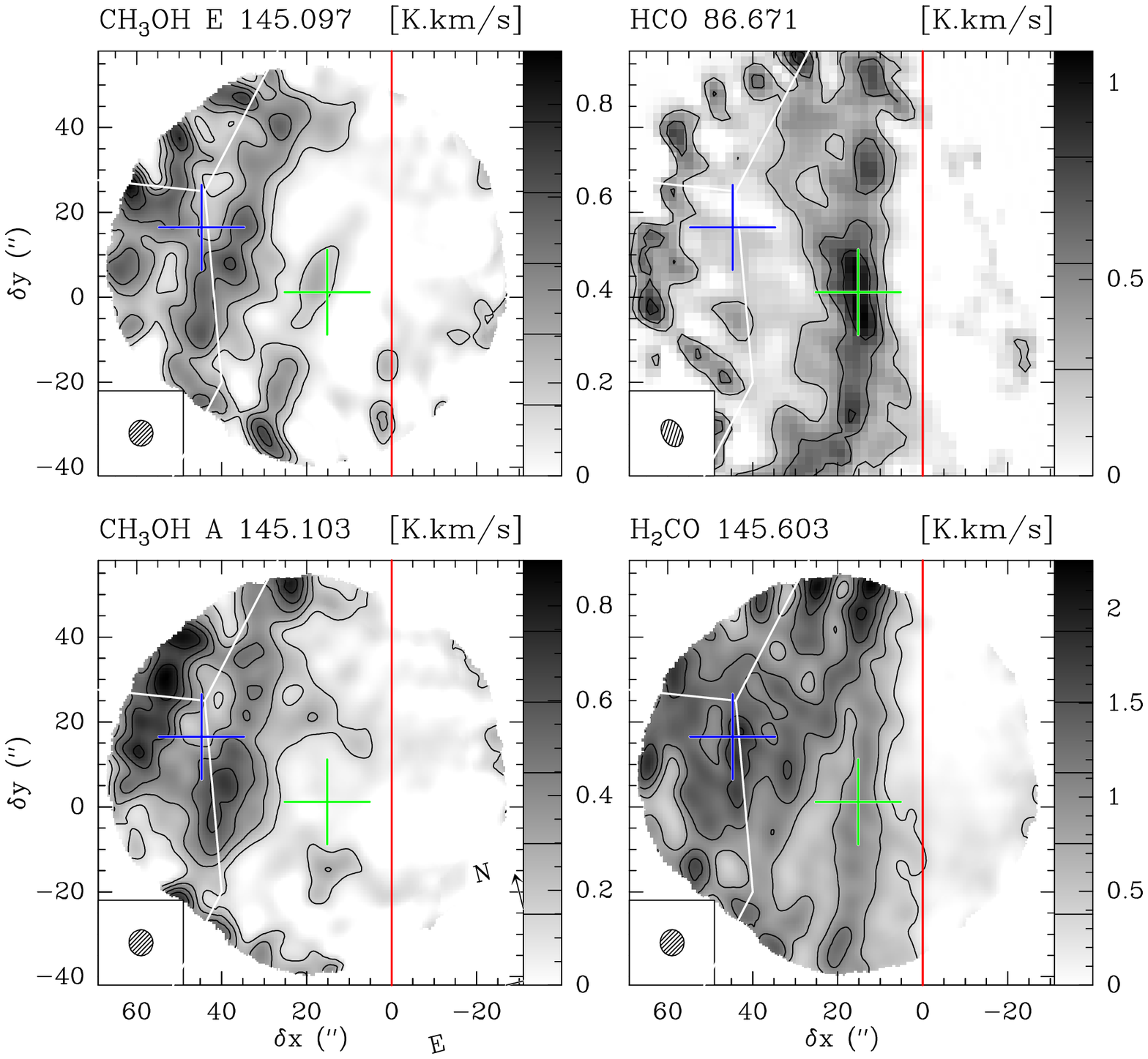}
    \caption{IRAM-PdBI maps of the Horsehead edge. The field of view
      is smaller than in Figure~\ref{fig:maps30m}. The angular
        resolution is $6''.7\times4''.4$ for HCO and $6''.1\times5''.6$
        for \hhco{} and \chhhoh{}. All other descriptions are
      identical to Figure~\ref{fig:maps30m}.}
    \label{fig:mapsPdBI}
  \end{figure}
}


\section{Introduction}

\TabObsMaps{}

Formaldehyde (\hhco{}) and methanol (\chhhoh{}) are key species in the
synthesis of more complex organic molecules, like amino acids and
other prebiotic molecules \citep{bernstein02,munoz-caro02,garrod08},
that could eventually end up in proto-planetary disks, and hence in
new planetary systems. It is therefore of great interest to understand
how these molecules are formed. Both species have been detected in a
wide range of interstellar environments, with typical gas-phase
abundances relative to H$_2$ of $\sim10^{-7}$ in hot cores
\citep[\eg{},][]{sutton95,ceccarelli00} and $\sim10^{-9}$ in cold dark
clouds \cite[\eg{},][]{bergman11}. They are also observed in shocked
regions, caused by an impact of molecular outflows on the surrounding
molecular clouds \citep[\eg{},][]{sakai12,codella12,tafalla10}.  Being
slightly asymmetric rotors, \hhco{} and \chhhoh{} are good tracers of
the physical conditions in Galactic and extragalactic molecular
clouds \citep{mangum93,leurini04,mangum13}.

Unlike \hhco{}, which can be formed both in the gas-phase and on the
surface of dust grains, \chhhoh{} is thought to be formed entirely on
the surfaces of dust grains, because classical ion-neutral chemistry
in the gas-phase alone cannot account for the observed \chhhoh{}
abundances \citep{garrod06a,geppert06}. The mechanism to form \chhhoh{}
on ices is thought to be the successive hydrogenation of CO, forming
\hhco{} as an intermediate product,
\begin{equation}
  \mathrm{CO} \xrightarrow[\h]{E_b} \mathrm{HCO} \xrightarrow[\h]{} \hhco
  \xrightarrow[\h]{E_b} \mathrm{CH}_2\mathrm{OH},\mathrm{CH}_3\mathrm{O}
  \xrightarrow[\h]{} \chhhoh,
  \label{eq:path}
\end{equation}
where $E_b$ are activation energy barries
\citep{tielens97,watanabe02}. Indeed, laboratory experiments have
succeeded to efficiently form both \hhco{} and \chhhoh{} through this
mechanism \citep{watanabe04,fuchs09}. Moreover, observations by the
\textit{Infrared Space Observatory (ISO)} and \textit{Spitzer} have
shown that dust grains are covered by ice mantles in the cold
envelopes surrounding high-mass protostars \citep{gibb00,gibb04},
low-mass protostars
\citep{boogert08,pontoppidan08,oberg08,bottinelli10} and in isolated
dense cores \citep{boogert11}. These studies revealed that the ice
mantles consist mostly of H$_2$O, CO$_2$ and CO, with smaller amounts
of \chhhoh{}, CH$_4$, NH$_3$ and \hhco{}. HCO ice has not been
detected in the interstellar medium (ISM), probably because its
formation on the ices is slower than its subsequent hydrogenation to
form \hhco{}. Indeed, the reactions starting with CO and \hhco{} in the
hydrogenation path (\ref{eq:path}) have activation energy barriers
\citep{fuchs09}.

Once these species are formed on the ices, they can be desorbed into
the gas-phase either through thermal or non-thermal processes. Thermal
desorption dominates in regions where dust grains reach temperatures
of at least 45~K for formaldehyde \citep{tielens87}, and 80~K for
methanol \citep{brown07,green09}. This happens in hot cores and hot
corinos \citep[\eg{},][]{jorgensen05,bisschop07,bottinelli07}, but
also in highly far-UV illuminated PDRs, like in the Orion bar
\citep{leurini10}. Non-thermal desorption by far UV photons can also
be efficient, as shown by laboratory
experiments~\citep{oberg09b,oberg09c}. Non-thermal desorption
dominates in colder regions, either UV-shielded dense cores where
secondary UV photons are produced by the interaction between cosmic
rays and H$_2$ molecules~\citep[\eg{},][]{caselli12} or in low
UV-field illumination photo-dissociation regions, where dust grains
are too cold to sublimate their ices. This is the case of the
Horsehead, where the combination of moderate radiation field
\citep[$\chi=60$ relative to the Draine field;][]{draine78}, and high
density ($\nH\sim10^4-10^5\pccm$) implies low dust grain temperatures,
from $\Td \sim 30$~K in the outer PDR to $\Td \sim 20$~K slightly
deeper inside the cloud~\citep{goicoechea09a}. \citet{guzman11}
compared single-dish observations of \hhco{} with PDR models including
grain-surface chemistry. They showed that the observed \hhco{}
abundance in the UV-illuminated edge of the Horsehead nebula can only
be explained by the formation of \hhco{} on the grains followed by its
photo-desorption into the gas-phase. Pure gas-phase chemistry was
enough to explain the \hhco{} abundance in the colder and UV-shielded
gas. The assignments of different formation routes were
strengthened by the different measured ortho-to-para ratio of
\hhco{}: the dense core displays an equilibrium value of $\sim3$,
while the PDR displays an out of equilibrium value of $\sim2$. The PDR
model predicted that \hhco{} will be one order of magnitude more
abundant in the PDR than the IRAM-30m beam-averaged measurements.

In order to further constrain the chemistry of complex organic
molecules and check the high abundance in the PDR position predicted
by the model, here we present high-resolution interferometric maps of
\hhco{} and \chhhoh{}, in addition to single-dish deep integrations of
the low lying rotational lines of \chhhoh{} towards two particular
positions in the Horsehead: the warm PDR and its associated cold dense
core. The observations and data reduction are presented in
Sect.~\ref{sec:obs}. The resulting single-dish and interferometric
data are described in Sect.~\ref{sec:obs-results}. In
Sect.~\ref{sec:abundances}, we compute the \chhhoh{} column densities
and abundances. We compare these results with PDR models in
Sect.~\ref{sec:models}. A discussion is presented in
Sect.~\ref{sec:discussion}. We summarize and conclude in
Sect.~\ref{sec:conclusions}.

\TabObsLines{}

\section{Observations and data reduction}
\label{sec:obs}

In this section, we describe the observations and data reduction of
the newly acquired \hhco{} and \chhhoh{} data. A detailed description
of the HCO, 218.222~GHz \hhco{}, $\dcop$ and 1.2mm continuum
observations and data reduction can be found in \cite{gerin09},
\cite{guzman11}, \cite{pety07} and \cite{hilyblant05}, respectively.
Tables~\ref{tab:obs:maps} and \ref{tab:obs:lines} summarize the
observation parameters of the data obtained with the IRAM-30m, PdBI
and CSO telescopes.

\subsection{PdBI maps}

We used the Plateau de Bure Interferometer (PdBI) to obtain $6''$
resolution maps of the emission of the \hhco{}
$2_{02}-1_{01}$ line at 146.603~GHz, the \chhhohE{}
$3_{-1}-2_{-1}$ line at 145.097~GHz and the \chhhohA{}
$3_0-2_0$ line at 145.103~GHz. These observations were
carried out in August and October 2011 with 5 antennas in the D
configuration and 6 antennas in the C configuration. The baseline
lengths ranged between 24 and 176~m. We observed a 19-field mosaic in
a hexagonal pattern covering an almost circular field-of-view of
$80''$ in diameter. The observations used about 19 hours of telescope
time. The on-source time scaled to a 6 antenna array is 5.3 hours,
after filtering out low quality visibilities. Two correlator windows
of 20~MHz (yielding a spectral resolution of 39~kHz) were concatenated
to cover both \chhhoh{} lines.  Another 20~MHz window was centered on
the targeted \hhco{} line. During the observations, the typical
precipitable water vapor amounts to 6~mm and the typical system
temperature was 145~K. The median noise level achieved over the mosaic
is 0.24~K and 0.12~K (\Tmb{}, in channels of 0.2\kms{} width) for
\hhco{} and \chhhoh{}, respectively.

We used the standard algorithms implemented in the \GILDAS{}\footnote{See
  \texttt{http://www.iram.fr/IRAMFR/GILDAS} for more information about the
  \GILDAS{} softwares~\citep{pety05}.}/\CLIC{} software to calibrate the PdBI
data. The radio-frequency bandpass was calibrated by observing the bright
quasar 3C454.3. Phase and amplitude temporal variations where calibrated by
fitting spline polynomials through regular mesurements of two nearby
quasars (0420$-$014 and 0528$+$134).  The PdBI secondary flux calibrator
MWC349 was observed once during every track, which allowed us to derive the
flux scale of the interferometric data. The absolute flux accuracy is
$\sim 10\%$.

\subsection{30m maps}

\FigLines{}

A multiplicative interferometer filters out the low spatial
frequencies, \ie{}, spatially extended emission. We thus observed the
same region with the IRAM-30m single dish telescope in order to
recover the low spatial frequency (``short- and zero-spacing'')
information filtered out by the PdBI. The \hhco{} $2_{02}-1_{01}$
(145.603~GHz) line was observed simultaneously with the \chhhohE{}
$3_{-1}-2_{-1}$ (145.097~GHz) and \chhhohA{} $3_0-2_0$ (145.103~GHz)
lines during $\sim13$ hours of average summer weather in August and
September 2012. We used both polarizations of the EMIR receivers and
the FTS backends at 49~kHz spectral resolution. We used the
position-switched, on-the-fly observing mode. The off-position offsets
were ($\delta$RA$,\delta$Dec) = $(100'', 0'')$, \ie{}, the
{\sc{H\,ii}} region ionized by $\sigma$Ori and free of molecular
emission. We completed the observations of the 146.097~GHz and
146.103~GHz \chhhoh{} lines with older but higher quality observations
obtained in January 2007. We used one polarization of the C150
receiver and one 20~MHz window of the VESPA correlator, yielding a
spectral resolution of 20~kHz. In this case, we used the
frequency-switched, on-the-fly observing mode, with a frequency throw
of 7.9~MHz. In both observations, we observed along and perpendicular
to the direction of the exciting star in zigzags (\ie{}, $\pm$ the
lambda and beta scanning direction). From our knowledge of the
IRAM-30m telescope, we estimate the absolute position accuracy to be
$3''$.

\FigLevelDiag{}

The IRAM-30m data processing was made with the \GILDAS{}/\CLASS{}
software. The data were first calibrated to the \Tas{} scale using the
chopper-wheel method \citep{penzias73}. The mixer tuning tables were
wrong for the frequency tuned in 2012. This basically turned the
single sideband mixers into double side band ones. As the actual
rejection was unknown, we could not recompute an accurate \Tas{}
scale. Instead, we fixed the 2012 \Tas{} scale according to the 2007
scale which was correct. To do this, we correlated the 2007 and 2012
\chhhoh{} data sets, which allowed us to determine the multiplicative
factors (1.9 and 2.4 for the vertical and horizontal polarizations,
respectively) needed to fix the 2012 data. We applied the same factors
for the \hhco{} data that were acquired with the same tuning. The data
was converted to main-beam temperatures (\Tmb{}) using the forward and
main-beam efficiencies (\Feff{} and \Beff{}). The resulting amplitude
accuracy is 10\%. We then computed the experimental noise by
subtracting a zeroth order baseline from every spectra. A systematic
comparison of this noise value with the theoretical noise computed
from the system temperature, the integration time, and the channel
width, allowed us to filter out outlier spectra. The spectra where
then gridded to a data cube through a convolution with a Gaussian
kernel.  Finally, we fitted another baseline of order 3 through each
spectra of the cube.

\subsection{Joint imaging and deconvolution of the interferometric and
  single-dish data}

\FigMapsSingleDish{}

Following~\citet{rodriguez08}, the \GILDAS{}/\MAPPING{} software and the
single-dish map from the IRAM-30m were used to create the short-spacing
visibilities not sampled by the Plateau de Bure interferometer. In short,
the maps were deconvolved from the IRAM-30m beam in the Fourier plane
before multiplication by the PdBI primary beam in the image plane. After a
last Fourier transform, pseudo-visibilities were sampled between 0 and 15~m
(the diameter of the PdBI antenna). These visibilities were then merged
with the interferometric observations. Each mosaic field was imaged and a
dirty mosaic was built combining those fields in the following optimal way
in terms of signal--to--noise ratio~\citep{pety10}. The resulting data
cubes were then scaled from Jy/beam to \Tmb{} temperature scale using the
synthesized beam size (see Table~\ref{tab:obs:maps}).

\subsection{Deep pointed integrations with the 30m}

These observations are part of the Horsehead WHISPER project (Wideband
High-resolution Iram-30m Surveys at two Positions with Emir Receivers,
PI: J. Pety), a complete and unbiased line survey at 1, 2 and 3~mm,
carried out with the IRAM-30m telescope. Two positions were observed:
1) the HCO peak, which is characteristic of the photo-dissociation
region at the surface of the Horsehead nebula \citep{gerin09}, and 2)
the $\dcop$ peak, that belongs to a cold condensation located less
than $40''$ away from the PDR edge, where HCO$^+$ and other species are
highly deuterated \citep{pety07}. Hereafter we will refer to these two
positions as the PDR and dense core, respectively. The combination of
the new EMIR receivers at the IRAM-30m telescope and the Fourier
Transform Spectrometers (FTS) yields a spectral survey with
unprecedented combination of bandwidth (36~GHz at 3mm, 25~GHz at 2mm
and 76~GHz at 1mm), spectral resolution (49~kHz at 3 and 2mm; and
195~kHz at 1mm), and sensitivity (median noise 8.1~mK, 18.5~mK and
8.3~mK, respectively). A detailed presentation of the observing
strategy and data reduction process will be given in a forthcoming
paper. In short, any frequency was observed with two different
frequency tunings and the Horsehead PDR and dense core positions were
alternatively observed every 15 minutes in position switching mode
with a common fixed off position. This observing strategy allows us to
remove potential ghost lines incompletely rejected from a strong line
in the image side band (the typical rejection of the EMIR sideband
separating mixers is only 13dB).

\subsection{Deep pointed integrations with the CSO}

We also report the detection of a new \hhco{} line in the
Horsehead. The \hhco{} $4_{13}-3_{12}$ at line at
300.836~GHz was observed at the Caltech Submillimeter Observatory
(CSO) telescope in October 2012. We used the Barney receiver and the
FFTS2 backends at 269~kHz spectral resolution. We used the
position-switched mode, with the same off-position used in the 30m
observations, \ie{} ($\delta$RA$,\delta$Dec) = $(100'', 0'')$. The
data processing was made with the \GILDAS{}/\CLASS{} software. A
fourth-order baseline was fit and subtracted from the spectra. The
data was converted to \Tmb{} using a beam efficiency of 0.68, that was
measured towards Jupiter.

\section{Observational results}
\label{sec:obs-results}

\subsection{Deep pointed integrations}

Figure~\ref{fig:lines} presents the \chhhoh{} lines detected in the
Horsehead. A diagram with the lower energy rotational levels of E-type
and A-type methanol is shown in Fig.~\ref{fig:level_diag}, with color
arrows indicating the detected lines. We detected 9 and 16 lines at
the PDR and dense core position, respectively. Within these lines, at
each position, 3 arise from A-type methanol and the rest arise from
E-type methanol. The brightest \chhhoh{} lines are the ones arising
from A-type symmetry species. All lines are brighter in the dense core
than in the PDR. The same behavior was found for \hhco{}
\citep{guzman11}. Gaussian fits of the \chhhoh{} lines at the PDR, in
general, result in broader line widths than at the dense
core. Although, a few lines have similar linewidths towards both
positions.

\FigMapsPdBI{}

\subsection{30m maps}

Figure~\ref{fig:maps30m} displays the single-dish maps of the 1.2mm
dust continuum emission, and the line integrated emission of the
$\dcop (3-2)$ (216~GHz) line, the \phhco{} $3_{03}-2_{02}$ (218~GHz)
and $2_{02}-1_{01}$ (146~GHz) lines, the \chhhohE{} $3_{-1}-2_{-1}$
(146~GHz) line and the \chhhohA{} $3_0-2_0$ (146~GHz) line. The
formaldehyde and methanol emission resembles, to first order, the dust
continuum spatial distribution. It first follows the
photo-dissociation front, \ie{} the top of the Horsehead nebula from
its front to its mane. It then forms two filaments almost
perpendicular to the photo-dissociation front, one of them delineating
the Horsehead throat. The E and A-type methanol emission shows similar
spatial distributions.

In contrast to the \dcop{} emission, which delineates a narrow filament,
formaldehyde and methanol emission is extended. The impression that
the formaldehyde emission is slightly more extended than the methanol
emission is an artifact related to the different signal-to-noise ratio of
these maps. Indeed, the 146~GHz E-type methanol map has by far the
lowest signal-to-noise ratio. The highest signal-to-noise ratio map
(\ie{}, the 146~GHz formaldehyde one) thus indicates that both
formaldehyde and methanol emit in the PDR region (green cross).

Although the methanol emission, in general, correlates well with the
formaldehyde emission, there are a few differences. First,
formaldehyde peaks on the right side (closest to the
photo-dissociation front) of the dense core as traced by the \dcop{}
emission, while methanol peaks on its left side. Second, the 146~GHz
formaldehyde line peaks at the $\dcop$ emission peak (blue cross),
where the gas is cold ($\Tkin\simeq20$~K) and dense
($\nH\sim10^5\pccm$), while \chhhoh{} presents a local minimum at the
same position.

\subsection{PdBI maps}

Figure~\ref{fig:mapsPdBI} displays the HCO $1_{01} 3/2,2 - 0_{00}
1/2,1$ (86.670~GHz), \hhco{} $2_{02}-1_{01}$ (145.602~GHz), \chhhohE{}
$3_{-1}-2_{-1}$ (145.097~GHz) and \chhhohA{} $3_0-2_0$ (145.103~GHz)
integrated intensity maps obtained with the PdBI. Contrary to the case
of HCO, which peaks in the PDR, \hhco{} and \chhhoh{} lines are
brighter in the more UV-shielded layers of the nebula. However, the
filament traced by the HCO emission is clearly seen in the \hhco{} map
at the PDR edge. The exact spatial distribution of the \chhhoh{}
emission near the PDR position is difficult to infer due to the low
signal-to-noise ratio at this position.

The minimum seen in the methanol 30m maps near the dense core position
(blue cross) is present in both methanol lines and it is preserved in
the higher angular resolution map. On the other hand, \hhco{}, which
was observed simultaneously with \chhhoh{}, peaks at the dense
core. This suggests that 1) the methanol gap is real and not an
artifact of the deconvolution, and 2) methanol and formaldehyde
emission show opposite behavior at the \dcop{} peak.

In contrast with HCO data, the maps of \hhco{} and \chhhoh{} emission
obtained with the PdBI only data (i.e., without the short-spacings
from the IRAM-30m telescope) looks like noise with a typical noise
level of 0.15~K. Indeed, the spectra extracted from the hybrid
synthesis (PdBI + 30m) cube at the position of the dense core and of
the PDR have an integrated area compatible, within the noise level,
with the 30m spectra at the same positions. This implies that the beam
dilution of the single-dish observations of H$_2$CO and CH$_3$OH is
marginal. We will thus use the higher signal-to-noise ratio 30m
spectra to infer the column densities at the PDR and dense core
positions.

\FigCSO{}

\section{Column densities and abundances}
\label{sec:abundances}

\subsection{\hhco{}}

Figure~\ref{fig:cso} shows the new $\ohhco$ $4_{13}-3_{12}$ line
detected in the Horsehead. \cite{guzman11} reported the detection of
four $\ohhco$ lines at lower frequencies. They used a nonlocal non-LTE
radiative transfer code from \cite{goicoechea06} adapted to the
Horsehead geometry to model the observed \hhco{} line intensities. The
parameters they used are $\Tkin=60$~K and $n(\hh) = 6\times10^4 \pccm$
at the PDR position and $\Tkin=20$~K and $n(\hh) = 10^5 \pccm$ at the
dense core position. The column density that best reproduced the
observations is given in Table~\ref{tab:column_densities} and the
modelled line profile is shown in red in Fig.~\ref{fig:cso}. The line
profiles for two other models are also shown. The new detected \hhco{}
line is in agreement with the predictions of our previous model, and
thus corroborates the \hhco{} column density derived by
\cite{guzman11}.

\subsection{\chhhoh{}}

Typical densities in the Horsehead ($\nH = 10^4-10^5 \pccm$) are lower
than the critical densities of the observed transitions of methanol
(see Table~\ref{tab:ncrit}). We therefore expect the lines to be
sub-thermally excited ($\Tex \ll \Tkin$), and a non-LTE approach is
needed to compute the \chhhoh{} column densities. E and A symmetries
of \chhhoh{} are treated as different species because radiative
transitions between them occur in timescales too long compared to the
lifetime of the molecules. We do not correct for beam dilution
factors because the interferometric maps indicate that beam dilution
is marginal for these molecules. The spectroscopic parameters
for the detected transitions are given in Table~\ref{tab:spec_param}.

\TabCritDens{}
\SpecParam{} 

We performed non-LTE radiative transfer models using the RADEX LVG
model \citep{vandertak07}, which computes the line intensities of a
species for a given column density, kinetic temperature and density of
$\hh$.  We included 100 rotational levels for $\chhhohE$ and
$\chhhohA$, where the maximum energy level lies at $\sim
155$~cm$^{-1}$ for both species. We considered $\phh$ and $\ohh$ as
collision partners with collisional excitation rates from
\cite{rabli10}. An \hh{} ortho-to-para ratio of 3 (high temperature
limit) was assumed in the models. Lower \hh{} ortho-to-para ratios
were also tested, and we find that the difference is negligible. We
also investigated the importance of electrons in the excitation at the
PDR. The electron fraction is $10^{-4}$ relative to H$_2$ in the PDR
position, while in the dense core the electron fraction is
$\sim10^{-9}$ \citep{goicoechea09b}. We have computed the
$\chhhohE-$electron and $\chhhohA-$electron collisional coefficients
within the dipolar Born approximation
\citep[\eg{},][]{itikawa71}. Owing to the relatively large dipole of
methanol (1.7~D), dipole-allowed cross sections are expected to be
dominant and mostly determined by the long-range electron-dipole
interaction. Thus, for the water molecule which has a similar dipole
(1.8~D), \citet{faure04} have shown that the Born approximation is
accurate down to typically 1~eV. At lower energy, short-range forces
can become important and these effects were found to reduce the low
energy cross sections by up to a factor of $\sim1.5$. In the case of
methanol, the Born treatment could overestimate the rotational cross
sections by a factor of $2-3$, as suggested by the measurements of the
total elastic cross section \citep[see][and references
  therein]{vinodkumar13}. In the Born approximation, cross sections
are proportional to line strengths and the square of the dipole and
therefore strictly obey the dipolar selection rule. In this work, line
strengths and dipoles were taken from the JPL catalog
\citep{Pickett98}. Excitation cross sections were computed in the
energy range 0.1~meV$-1$~eV and rate coefficients were deduced in the
range $10-1000$~K, for the lowest 256 levels of \chhhohA{} (1853
transitions) and the lowest 256 levels of \chhhohE{} (2324
transitions).

We first constrained the column density of $\chhhohE$, as we detected
more lines of this species. The density profile across the PDR in the
Horsehead is well constrained \citep{habart05}. Several efforts have
been made to compute the thermal profile, but it remains less
constrained than the density. We therefore decided to fix the $\hh$
density and vary the temperature in our models. For this, we run grids
of models for kinetic temperatures of $10-100$~K and $\chhhohE$ column
densities between $10^{11}$ and $10^{15} \pscm$. The density of $\hh$
was kept constant to the well-known values of $\nH = 6\times10^{4}
\pccm$ in the PDR, and $\nH = 1\times10^5 \pccm$ in the dense
core. Then, the $\chi^2$ was computed\footnote{$\chi^2 =
  \sum\limits_{1}^N
  \frac{(\mathrm{Observation}-\mathrm{Model})^2}{\sigma^2}$, with $N$
  the number of detected lines.} at each point of the grid to
determine the best fit parameters. Figure~\ref{fig:chi2} shows the 1,
2 and 3$\sigma$ confidence levels for the PDR (red) and dense core
(blue) for models without electron excitation. The best fits are for
kinetic temperatures of 60~K (PDR) and 30~K (core), in good agreement
with previous determinations. The $\chhhohE$ column densities are well
constrained. The best fits are for $\chhhohE$ column densities of
$(2.7\pm0.5)\times10^{12} \pscm$ and $(6.5\pm0.8)\times10^{12} \pscm$
for the PDR and dense core, respectively. We also run models for a
fixed kinetic temperature instead of a fixed gas density. These models
result in similar \chhhoh{} column densities, but in a $\hh$ gas
density of $\nH=6\times10^4\pccm$ at the dense core position, which is
lower than expected ($\nH\sim10^5\pccm$). 

We find that the change in the $\chhhohE$ column density when
including the electron excitation is negligible in these
models. However, for a grid of models with the three parameters
($\Tkin, \nH$ and $N(\chhhoh)$) free and no electron excitation, we
find a low kinetic temperature ($\Tkin=40$~K) and a high density
($\nH=10^5$) at the PDR. When including the electron excitation, on
the other hand, we obtain $\Tkin\sim65$~K and
$\nH\sim5\times10^4\pccm$, in much better agreement with previous
estimates. The inferred \chhhoh{} column density is the same for both
models ($N=2.7\times10^{12}\pscm$). Hence, collisions with electrons
are not important to determine the \chhhoh{} column density, but they
are important to determine the kinetic temperature and density of the
gas. The electron excitation was found to be important to determine
the column density of CH$_3$CN in the PDR \citep{gratier13}.


The determination of the $\chhhohA$ column density is more difficult
because only 3 lines were detected, providing less constrains to the
model. We thus fixed the kinetic temperature to the best fit values
found for $\chhhohE$, in addition to keeping the density of $\hh$
constant, and run models varying only the column density. The best
fits are for $\chhhohA$ column densities of $(2.0\pm0.3)\times10^{12}
\pscm$ and $(8.1\pm1.0)\times10^{12} \pscm$ for the PDR and
dense core, respectively.


Table~\ref{tab:column_densities} summarizes the derived column
densities and abundances relative to total number of atomic hydrogen
atoms for HCO, \hhco{} and \chhhoh{}. The E/A methanol ratio is
$1.3\pm0.3$ and $0.8\pm0.1$ for the PDR and dense core,
respectively. The total \chhhoh{} abundance is similar in the PDR and
dense core, being only a factor $\sim1.9$ larger in the dense
core. Similar abundances were also found for \hhco{} in both
positions. The $\hhco/\chhhoh$ abundance ratio is $\sim2.3\pm0.4$ in
the PDR and $\sim0.9\pm0.1$ in the dense core.


\FigChi{}
\TabColumnDens{}
\FigPDRmodel{}
\subsection{Deuterated \chhhoh{}}

\cite{guzman11} detected single and doubly deuterated formaldehyde in
the Horsehead and derived [HDCO]/[\hhco{}] = 0.11 and
[D$_2$CO]/[\hhco{}] = 0.04 in the dense core. We now searched for
deuterated methanol at this position. We did not detect any
transitions from deuterated methanol, but we computed an upper limit
for the CH$_2$DOH column density toward the dense core. The low energy
transitions $2_K - 2_K$ of CH$_2$DOH lie at 89~GHz. The RMS noise at
this frequency is 6.43~mK with 0.17$\kms$ spectral
resolution. Assuming a linewidth of 0.5$\kms$, we obtain an upper
limit for the integrated line intensity of $6\mKkms$. This translates
into a 1$\sigma$ upper limit for the column density of
$1.4\times10^{12}\pscm$ and a fractionation ratio
[CH$_2$DOH/$\chhhoh]\leq0.26$ at the dense core position. Assuming
that methanol has fractionation levels similar to that of
\hhco{}, we conclude that our data is not sensitive enough to detect
lines from deuterated methanol molecules. For reference, singly,
doubly and even triply deuterated methanol species, such as CH$_2$DOH,
CHD$_2$OH and CD$_3$OH have been detected toward low-mass protostars
with abundance ratios of $37-65$\% (CH$_2$DOH), 20\% (CHD$_2$OH) and
1.4\% (CD$_3$OH) compared to their non-deuterated
isotopologue \citep{parise04,parise06}.

\section{Comparison with models}
\label{sec:models}

In this section we study the HCO, \hhco{} and \chhhoh{} chemistry in
the Horsehead. We used an updated version of the one-dimensional,
steady-state photochemical model \citep{lepetit06}, used in the study
of \hhco{} by \cite{guzman11}. We include the density profile
displayed in the upper panel of Fig.~\ref{fig:meudon}, a radiation
field $\chi = 60$ \citep[relative to the Draine field;][]{draine78},
the elemental gas-phase abundances from \cite{goicoechea09b} (see
their Table 5) and a cosmic ray ionization rate $\zeta = 5 \times
10^{-17} \ps$ per $\hh$ molecule.

We compare the results of a pure gas-phase chemical network with one
that also includes grain surface reactions. Fig.~\ref{fig:meudon}
displays the results for the photochemical models. The left column
presents the results computed by the code, which samples the
UV-illuminated gas on a finer spatial grid than the UV-shielded gas to
correctly represent the steep physical and chemical gradients. The
right column presents the results convolved with a Gaussian of 6''
(FWHM) to ease the comparison with the abundances inferred from the
PdBI observations at 6'' angular resolution. The predicted \hhco{} and
\chhhoh{} abundance profiles are shown in the middle panel. The
$\chhhoh/\hhco$ abundance ratio is shown in the lower panel. The
horizontal bars show the results inferred from observations. Results
for the pure gas-phase model are shown in dashed lines, and results
for the model including grain surface chemistry are shown in solid
lines.

\subsection{Pure gas-phase models}
\label{sec:gas}

We used a modified version of the pure gas-phase chemical network of
the \textit{Ohio State University (OSU)}. Our network includes the
photo-dissociation of HCO and of \hhco{} (leading to CO and $\hh$),
and the atomic oxygen reaction with the methylene radical (CH$_2$) to
explain the high abundance of HCO in the PDR \citep{gerin09}. The
\hhco{} photo-dissociation channel that leads to HCO + H is also
included. 

As already shown by \cite{guzman11}, the pure-gas phase model
satisfactorily reproduces the observed \hhco{} abundance in the
dense-core ($\delta x \sim 35''$), but it underestimates the abundance
in the PDR ($\delta x \sim 15''$) by at least one order of
magnitude. In this model the formation of \hhco{} is dominated by
reactions between oxygen atoms and the methyl radical (CH$_3$) in both
the PDR and dense core. The destruction of \hhco{} in the PDR is
dominated by reactions with S$^+$ and by photo-dissociation, while it
is dominated by reactions with ions (mainly H$_3$O$^+$, HCO$^+$ and
S$^+$) in the dense-core.

In the gas-phase, methanol is mainly produced by the dissociative
recombination of CH$_3$OH$^+_2$, which is formed by reactions between
CH$^+_3$ and water. It is now well established that this reaction is
not efficient enough to explain the observed \chhhoh{} abundances
\citep{garrod06a,geppert06}. Indeed, the pure gas-phase model
underestimates the \chhhoh{} abundance by $\sim5$ orders of magnitudes
in both the PDR and dense core. Note that the \chhhoh{} abundance
predicted by the pure gas-phase model has been multiplied by 1000 in
Fig.~\ref{fig:meudon}.

\subsection{Grain surface chemistry models}
\label{sec:grain-chem}

We included the reactions on the surface of dust grains to the
gas-phase chemical network described in section~\ref{sec:gas}. A
detailed description of the implementation of grain surface chemistry
in the Meudon PDR code will be given in a subsequent paper
(Le Bourlot et al., to be submitted). The model includes the
adsorption, desorption and diffusive reactions on grains, where the
sequence to form formaldehyde and methanol is the one shown in the
hydrogenation path (\ref{eq:path}). We also introduce water formation
via hydrogenation reactions of O, OH until H$_2$O.
The hydrogenation of formaldehyde can lead to $\chhho$ and/or
CH$_2$OH. Gas-phase methoxy radical ($\chhho$) was recently detected
in the cold dense core B1-b \citep{cernicharo12}. We do not
distinguish between these two isomers, \ie{} $\chhho$ and CH$_2$OH, in
our network. The energy barriers associated with the H+CO (200~K) and
H+\hhco{} (300~K) reactions have been modified with respect to
\cite{guzman11}. As shown by laboratory studies
\citep{oberg09b,oberg09a,munoz-caro10}, photo-desorption can be an
efficient mechanism to release molecules into the gas phase in regions
exposed to radiation fields. We included photodesorption yields of
$10^{-3}$ and $2\times10^{-4}$ molecules per incident UV photon, for
\hhco{} and \chhhoh{}, respectively. These values are close to the
ones measured in the laboratory for CO, CO$_2$, H$_2$O and \chhhoh{}
\citep{oberg07,oberg09a,oberg09b,oberg09c}. Photodesorption of
adsorbed \chhhoh{} can produce both methanol and formaldehyde in the
gas-phase with a 50/50 branching ratio. This channel was not included
in the model of \citet{guzman11}, where photodesorption of adsorbed
\chhhoh{} produced only gas-phase methanol. We checked that for the
physical conditions prevailing in the Horsehead PDR and dense core,
the use of steady state chemistry in the model is a valid
assumption. Indeed, the formation/destruction timescales of \hhco{}
and \chhhoh{} are $<30$~yr and $<5000$~yr at the PDR and dense core,
respectively, while the estimated age of the Horsehead is
$\sim5\times10^5$~yr \citep{pound03}.

The new results are shown as solid lines in Fig.~\ref{fig:meudon}. The
dust temperatures in this model are 20~K. The model predicts two
abundance peaks at the edge of the nebula ($\delta x\sim15$ and
$\delta x\sim20$), with \chhhoh{} peaking deeper inside the cloud than
HCO and \hhco{}, and an abundance plateau at the inner layers of the
cloud. This way, HCO and \hhco{} dominate at the outer layers of the
cloud, while \chhhoh{} dominates in the inner layers. At the dense
core position ($\delta x\sim35''$), the modeled \hhco{} abundance is
similar to the one predicted by the pure gas-phase model. The
\chhhoh{} abundance, on the other hand, increases by $\sim5$ orders of
magnitude with respect to the pure gas-phase model. The observed
\hhco{} and \chhhoh{} abundances in the dense core are well reproduced
by the model including both gas-phase and grain surface chemistry. In
the PDR, this model is also in good agreement with the observed HCO
and \chhhoh{} abundances, but the predicted \hhco{} abundance is
higher than what is observed.
  
In this model, \chhhoh{} is produced by photodesorption of \chhhoh{}
ices everywhere in the cloud. At the dense core, the FUV photons that
photodesorb the \chhhoh{} ices are secondary photons produced by the
interaction between cosmic rays and \hh{} molecules. Gas-phase
\chhhoh{} is mainly destroyed by photodissociation in the PDR, and by
freeze-out in the dense core. \hhco{} is produced mainly by direct
photodesorption of \hhco{} ices in the PDR position. In the model of
\cite{guzman11}, we found that the dominant formation path for \hhco{}
in the dense core was the gas-phase route involving CH$_3$ and O. With
this new model, we find that at the dense core, \hhco{} is produced
mainly by photodissociation of \chhhoh{} ices. This difference is the
result of the several modifications made to the model, where the most
important one is the inclusion of the channel leading to \hhco{} in
the photodesorption of \chhhoh{} ice. \hhco{} is mainly destroyed by
photodissociation in the PDR and by reactions with ions (mainly
N$_2$H$^+$ and H$^+_3$) in the dense core\footnote{We first run models
  assuming the solar sulfur abundance (S/H$=1.38\times10^{-5}$). With
  this S abundance and including the grain surface chemistry, the
  destruction of \hhco{} is dominated by reactions with S$^+$ in both
  the PDR and dense core. We then run models using a lower sulfur
  abundance (S/H=$3.5\times10^{-6}$), which is the one that reproduces
  the observed CS and HCS$^+$ abundances \citep{goicoechea06}. In this
  model, which is the one adopted in this paper, S$^+$ is not
  important in the destruction of \hhco{}. As a result, the predicted
  \hhco{} abundance at the dense core is 3 times higher in the low-S
  abundance model, and the agreement between model and observations is
  better. We conclude that a sulfur abundance of
  S/H$=3.5\times10^{-6}$ can reproduce the observed abundance of CS,
  HCS$^+$ and \hhco{}.}

It is important to remember that models including complex grain
surface processes use numerous physical parameters, such as the
adsorption energy, diffusion energy of absorbed species, the
activation energy of the different hydrogenation reactions and the
desorption yield, which have large uncertainties. The results
predicted by these models are highly dependent on the values used for
these parameters. Providing observational results is key to constrain
these parameters and benchmark the models.

\section{Discussion}
\label{sec:discussion}

\subsection{Observational evidence of different \hhco{} formation
  mechanisms} 

The observations suggest that at the PDR, \hhco{} is formed on the
surface of dust grains, and at the dense core, \hhco{} is mostly
formed in the gas-phase. First, we find a different ortho-to-para
ratio in each position: the equilibrium value of 3 is found at the
dense core, while a lower out-of-equilibrium value of 2 is found at
the PDR. Second, if \hhco{} is formed on the grains, its spatial
distribution should resemble the spatial distribution of
\chhhoh{}. However, the high-resolution PdBI maps show that the
\chhhoh{} emission presents a minimum at the dense core position,
while \hhco{} peaks at this position. The easiest interpretation is
that 1) at the dense core, the photodesorption of both \hhco{} and
\chhhoh{} is not efficient, so the ices remain depleted onto grains,
and that 2) \hhco{} is formed on the gas-phase at this
position. Indeed, the pure gas-phase model reproduces well the \hhco{}
abundance at the dense core. Moreover, the number of photons that can
photodesorb species at the dense core are much less than the number of
photons available at the PDR position. This way, gas-phase \chhhoh{}
is present in an envelope around the dense core, while \hhco{} is
present in both the envelope and the dense core itself. This is
consistent with the lower gas density inferred from the radiative
transfer analysis of \chhhoh{} at the dense core position.

Depletion of \chhhoh{} has also been observed in other dense cores.
\cite{tafalla06} found that \chhhoh{} shows a dip in its emission at
the center of the dense cores L1498 and L1517B. They concluded that
the drop in emission corresponds to a real drop in column density,
suggesting that \chhhoh{} freeze out onto grains at the center of
these dense cores.

\subsection{Comparison with other environments}

\hhco{} and \chhhoh{} have been detected in the Orion Bar, which is a
more extreme case of a PDR (higher $\chi$, $\Tkin$ and
$\Td$). \cite{leurini10} estimated an abundance ratio $\hhco/\chhhoh =
14-1400$ in the interclump medium ($\nH\sim10^4 \pccm$) and
$\hhco/\chhhoh = 0.9-2.5$ in the denser clumps ($\nH\sim10^6
\pccm$). Despite the large uncertainties and the fact that the clumpy
medium was not resolved, they concluded that the abundance of methanol
relative to formaldehyde decreases by at least one order of magnitude
in the interclump medium in comparison to the dense clumps. Unlike the
Orion Bar, we do not find a large difference in the $\hhco/\chhhoh$
abundance ratio between the PDR and dense core in the Horsehead
($\hhco/\chhhoh$ is only $\sim2$ times larger in the PDR).  This is
probably due to the difference in the radiation field between the two
sources ($\chi\simeq 10^4$ in the Orion Bar; $\chi=60$ in the
Horsehead). Hence, photodissociation of \chhhoh{} (that produces
\hhco{}) is more efficient in the Orion Bar, specially in the
interclump medium, and the $\hhco/\chhhoh$ is thus larger. The
$\hhco/\chhhoh$ abundance ratio we derive in the Horshead ($0.9-2.3$)
is very similar to the one found in the dense clumps of the Orion
Bar. These clumps are more protected from the FUV photons, and thus
resemble more the conditions prevailing in the Horsehead. Indeed,
$\chi/n\sim100/10^4$ in the Horsehead, which is similar to the ratio
found in the dense clumps of the Orion Bar ($\chi/n\sim10^4/10^6$). 


\cite{maret04,maret05} measured the formaldehyde and methanol
abundances towards a sample of low mass protostars. From these
studies, an abundance ratio $\hhco/\chhhoh=0.7-4.3$ is found for the
hot corino region. In the low mass starless cores, L1498 and L1517B,
\cite{tafalla06} found $\hhco/\chhhoh=1.1-2.2$. These values are
comparable to the abundance ratio derived in the Horsehead. A lower
abundance ratio is found in hot cores
\cite[$\hhco/\chhhoh=0.13-0.28$;][]{bisschop07} and an even lower
ratio is found in shocked gas in Galactic center clouds
\citep[$\hhco/\chhhoh\sim0.01$;][]{raquenatorres06}. In the diffuse
medium, \hhco{} is widely seen \cite[\eg{},][]{liszt06} but methanol
has not been detected, although it has been searched for by
\citep{liszt08}. They derive a upper limit for the \chhhoh{} column
density which translates into a lower limit for the abundance ratio of
$\hhco/\chhhoh \geq12$.

\subsection{Thermal and non-thermal desorption}

In hot cores, hot corinos and shocked regions the ices that were
formed in the cold gas in the prestellar stage are completely
sublimated into the gas-phase through thermal desorption and
sputtering. The observed \hhco{} and \chhhoh{} gas-phase abundances in
these regions should therefore resemble the original ice
composition. Observations of ices in the envelopes of low-mass
protostars show large variations of the \chhhoh{} ice abundance with
respect to water ice \citep[1\%-30\%;][]{boogert08}. In cold cores, a
methanol ice abundance of 5\%-12\% relative to water has been found
\citep{boogert11}. The variations in the \chhhoh{} ice abundance are
caused by the different local physical conditions, and thus reflect
the variations in the evolutionary stages of these sources
\citep{cuppen09}. The abundance of \hhco{} ice has been more difficult
to determine because it has been detected in just a few sources and
its stronger ice feature at $6~\mu$m is blended with that
of HCOOH \citep{oberg11}. A relatively constant \hhco{} ice abundance
of $\sim6$\% relative to water is found in low-mass protostars
\citep{gibb04,boogert08}. This value seems to be independent from the
\chhhoh{} ice abundance. From these detections, an ice abundance ratio
$\hhco/\chhhoh\simeq0.2-6.0$ is obtained in low-mass protostars, which
is consistent with the gas-phase abundance ratio found in hot corinos
by \citep{maret04,maret05}.

In cold cores and low-UV illumination PDRs, the ices are released
through non-thermal processes, like photodesorption. The
$\hhco/\chhhoh$ ratio will thus depend on the dust temperature, and on
the number of UV photons that will desorb the ices into the
gas-phase. The \chhhoh{} photodesorption yield has been previously
found to be $\sim10^{-3}$ molecules per incident UV photon, similar to
other species like CO, CO$_2$ and H$_2$O
\citep{oberg09a,oberg09b,oberg09c}. 
Although the \chhhoh{} photodesorption yield is uncertain, the one
adopted in our model gives a \chhhoh{} abundance that is consistent
with the observations.  As discussed in Section~\ref{sec:grain-chem},
the predictions of chemical models including grain surface chemistry
are highly dependent on the different physical parameters.  Future
laboratory studies, using different ice mixtures (similar to those
observed in the ISM), will help to better understand the non-thermal
desorption of ices into the gas-phase and will hopefully provide
precise values for the photodesorption yields and other parameters
needed in the models.



\section{Summary and conclusions}
\label{sec:conclusions}

We have presented deep observations of \chhhoh{} lines toward the
Horsehead PDR and its associated dense core, together with single-dish
and interferometric observations of \hhco{} and \chhhoh{}. In general,
formaldehyde and methanol emission is extended, although there are a
few differences. The \chhhoh{} emission presents a minimum at the
dense core position, that is also seen in the lower angular resolution
30m map, while \hhco{} peaks at this position. From the
high-resolution maps, we conclude that beam dilution of the
single-dish deep integrations is marginal. This is in contrast with
the model prediction of an abundance peak at the PDR position. We
therefore confirm the \hhco{} abundances inferred by \citet{guzman11}
and derive a similar \chhhoh{} abundance with respect to H
($\sim10^{-10}$) in the PDR and dense core position, that is also
similar to the observed abundance of \hhco{}. We find that collisions
with electrons are not important to determine the \chhhoh{} column
density when the physical conditions of the gas are known. However, in
order to determine the kinetic temperature and density of the gas,
electron excitation should be taken into account.  The inferred
methanol E/A ratio is close to one in both the PDR and dense core
position. The $\hhco/\chhhoh$ abundance ratio is $2.3\pm0.4$ in the
PDR and $0.9\pm0.1$ in the dense core.

At the PDR, observations suggest that both \hhco{} and \chhhoh{} are
formed on the surface of dust grains and are subsequently released
into the gas-phase through photodesorption. Indeed, pure gas-phase
chemical models predict \hhco{} and \chhhoh{} abundances that are too
low compared to what is inferred from the observations. At the dense
core, on the other hand, the dominant formation mechanism differs for
\hhco{} and \chhhoh{}. \hhco{} is mostly formed in the gas-phase,
while \chhhoh{} is formed on the grains. Indeed, a pure gas-phase
model can reproduce the observed \hhco{} abundance at this position,
while it fails by $\sim5$ orders of magnitude for \chhhoh{}. Moreover,
the high-resolution PdBI maps show that \chhhoh{} is present in an
envelope around the dense core, while \hhco{} is present in both the
envelope and the dense core itself. The different \hhco{} formation
mechanisms in the PDR and dense core are strengthened by the different
ortho-to-para ratio inferred from the observations ($o/p\sim2$ in the
PDR and the equilibrium value $o/p\sim3$ in the dense core).
 
\begin{acknowledgements}
  We thank the IRAM PdBI and 30~m staff for their support during the
  observations. We thank Simon Radford for his help during the CSO
  observations and data reduction. VG thanks support from the Chilean
  Government through the Becas Chile scholarship program. This work
  was also funded by grant ANR-09-BLAN-0231-01 from the French {\it
    Agence Nationale de la Recherche} as part of the SCHISM
  project. J.R.G. thanks the Spanish MINECO for funding support from
  grants AYA2012-32032, AYA2009-07304 and CSD2009-00038. J.R.G. is
  supported by a Ramon y Cajal research contract.
\end{acknowledgements}

\bibliographystyle{aa}
\bibliography{bibliography}


\end{document}